\documentclass[12pt,preprint]{aastex}  
\usepackage{emulateapj5}

\def\chandra{{\it Chandra}}  
\def\xmm{{\it XMM-Newton}}  
  
\def\hst{{\it HST}}

\def\merlin{{\it MERLIN}}  
\def\vla{{\it VLA}}

\def\lum{erg s$^{-1}$}  
\def\flux{erg cm$^{-2}$ s$^{-1}$}  
  
\def\arcsec{$^{\prime\prime}$}  
\def\deg{$^{\circ}$}  
  
\def\ltsima{$\; \buildrel < \over \sim \;$}  
\def\simlt{\lower.5ex\hbox{\ltsima}} 
\def\gtsima{$\; \buildrel > \over \sim \;$}  
\def\simgt{\lower.5ex\hbox{\gtsima}} 
  
\def\3c{3C~371}  
\def\pk{PKS~2201+044}

\begin{document}  
  
\title{Deep Chandra and multicolor HST observations \\  
of the jets of 3C~371 and PKS~2201+044}  
  
\author{Rita M. Sambruna, Davide Donato}  
\affil{NASA/GSFC, Code 661, Greenbelt, MD 20771  
(Rita.M.Sambruna@nasa.gov)}  
  
\author{F. Tavecchio and L. Maraschi}  
\affil{Osservatorio Astronomico di Brera, via Brera 28, 20121 Milano, Italy}  
  
\author{C.~C. Cheung\altaffilmark{1}}  
\altaffiltext{1}{Jansky Postdoctoral Fellow; National Radio Astronomy  
Observatory}  
\affil{Kavli Institute for Particle Astrophysics and Cosmology,  
Stanford University, Stanford, CA 94305}   
  
\author{C. Megan Urry}  
\affil{Yale University, New Haven, CT 06520}

\begin{abstract}  
  
This paper presents multiwavelength imaging and broad-band
spectroscopy of the relativistic jets in the two nearby radio galaxies
\3c\ and \pk, acquired with \chandra, \hst, \vla, and \merlin. Radio
polarization images are also available. The two sources stand out as
``intermediate'' between FRIs and FRIIs; their cores are classified as
BL Lacs, although broad and narrow optical emission lines were
detected at times. The multiwavelength images show jet morphologies with the
X-ray emission peaking closer to the nucleus than the longer
wavelengths. The jets are resolved at all wavelengths in a direction
perpendicular to the jet axis. The jets SEDs are consistent with a
single spectral component from radio to X-rays, interpreted as
synchrotron emission. The SEDs show a progressive softening from the
inner to the outer regions of the jet, indicating that the electron
break energy moves to lower energies with distance from the
core. Overall, the X-ray and multiwavelength properties of the jets of
\3c\ and \pk\ appear intermediate between those of FRIs and FRIIs.

\end{abstract}


\section{Introduction}  
  
Recent studies of extragalactic radio jets with \chandra\ and \hst\  
demonstrated the power of multiwavelength imaging spectroscopy to  
understand the physical properties and structure of these  
systems. These observations established that the X-ray morphology and  
emission processes of kpc-scale jets are closely related to radio  
morphology. In low-power Fanaroff-Riley I (FRIs; Fanaroff \& Riley  
1974) radio galaxies, the X-ray emission falls off more rapidly away  
from the core than does the longer wavelength emission; the jet   
radio-to-X-ray Spectral Energy Distributions (SEDs) can be described  
by single or broken power law spectra (Worrall et al. 2003),  
suggesting the X-ray emission from these jets is due to synchrotron  
radiation from the same electron population responsible for the  
radio.   
  
In higher-power FRIIs, the X-ray jet morphology often follows very
closely the radio one (Sambruna et al. 2002, 2004, 2006; Marshall et
al. 2005) and individual knots have complex SEDs with X-ray flux above
the extrapolation from the optical. In 3C~273, however, the connection
between optical and X-ray spectra from the jet knots is unclear
allowing different interpretations (Uchiyama et al. 2006). The origin
of the X-ray emission is still debated, ranging from inverse Compton
emission off the CMB (IC/CMB; Tavecchio et al. 2000; Celotti et
al. 2001) to synchrotron emission from an electron population with an
excess of high-energy particles resulting from acceleration (Stawarz et
al. 2004) or losses limited by Klein-Nishina effects (Dermer \& Atoyan
2002), to synchrotron emission from protons (Aharonian 2002). In some
sources, the X-ray jets peak upstream from the radio and optical
(Sambruna et al. 2006), suggesting deceleration by entrainment of
ambient gas in the context of IC/CMB (Georganopoulos \& Kazanas 2004;
Tavecchio et al. 2006). Thermal emission from shocked ambient gas hit
by the relativistic jet is also observed (Ly et al. 2005), allowing to
derive the gas density and the jet speed.
  
Interestingly, high-energy emission was also detected from the jets of
a few BL Lacertae sources, traditionally unified to FRIs (Wardle,
Moore, \& Angel 1984; Kollgaard et al. 1992). Early \chandra\
observations detected X-ray counterpart to the optical jet of
PKS~0521--365 (Birkinshaw et al. 2002), where the broad-band emission
was found to be consistent with synchrotron emission from a broken
power law populations of electrons, similar to FRIs. Another X-ray jet
was detected by us (Pesce et al. 2001, P01 in the following) in a
short (10~ks) ACIS-S exposure of 3C~371 aimed at studying the
circumnuclear environment (Donato, Sambruna, \& Gliozzi 2004). We
found evidence for a synchrotron origin of the X-ray emission from an
electron population with decreasing high-energy cutoff; however, the
ACIS-S exposure was too short to study the X-ray jet properties in
detail, in particular the X-ray morphology and spectra. An X-ray jet
is also detected in 2007+777 in our ACIS-S image (Sambruna et
al. 2007).
  
We later acquired deeper \chandra\ and multicolor HST observations of
\3c\ and \pk, which are the focus of this paper.  The two sources were
classified as BL Lacs on the basis of their non-thermal continuum
properties, although optical emission lines are detected (see
\S~2.1). In both cases, the core-jet radio morphology resembles that
of FRIIs; however, their radio power and the optical magnitude of
their hosts (Urry et al. 2000) firmly put them in the FRI region of
the Owen-Ledlow plane (Ledlow \& Owen 1996). As part of our program of
multiwavelength imaging-spectroscopy of kpc-scale jets, we obtained
deep \chandra\ and multicolor \hst\ observations of both \3c\ and
\pk. Here we present the X-ray and optical observations, as well as
new radio data.
  
The paper is structured as follows. Observations and sources  
properties are summarized in \S~2. The results for the cores and jets 
are given in \S\S~3 and 4, respectively, while discussion and  
conclusions follow in \S~5. Throughout this paper, a concordance  
cosmology with H$_0=71$ km s$^{-1}$ Mpc$^{-1}$,  
$\Omega_{\Lambda}$=0.73, and $\Omega_m$=0.27 (Spergel et al. 2003) is  
adopted. With this choice, 1\arcsec\ corresponds to 0.983~kpc for  
\3c\ and to 0.535~kpc for \pk. The energy spectral index, $\alpha$,   
is defined such that $F_{\nu} \propto \nu^{-\alpha}$, and the photon
index $\Gamma=\alpha+1$.

\section{Observations}

\subsection{The sources}   
  
\noindent{\bf 3C~371:} This is a nearby ($z$=0.051) radio source,  
optically classified as a BL Lac based on variability and polarization  
(Angel \& Stockman 1980). It was previously classified as an N galaxy  
(Miller 1975), based on the presence of narrow optical emission lines  
in the spectrum. Its radio morphology is more similar to an FRII, with  
two giant lobes and a 25\arcsec-long, one-sided jet (Wrobel \& Lind  
1990; Kollgaard et al. 1996). The jet-to-counterjet flux ratio in the  
radio indicates \3c\ is seen at moderate viewing angles, $\theta \le  
18$\deg\ (Gomez \& Marscher 2000).  
  
An optical counterpart to the radio jet was detected in ground-based  
images (Nilsson et al. 1997) and with \hst\ (Scarpa et al. 1999),  
while X-ray emission from the jet was discovered by us in an  
exploratory 10~ks \chandra\ observation (P01). The optical morphology  
of the jet is similar to the radio, with an inner weaker knot (knot B)  
at 1.9\arcsec\ from the core, a brighter one (knot A) at 3.1\arcsec\  
(Scarpa et al. 1999). The X-ray image indicated a different  
morphology, with the jet emitting most of the X-ray counts at knot B  
and fading toward the end of the jet.  In both knots, the X-ray flux  
lies below the extrapolation of the radio-to-optical continuum. We  
interpreted the SEDs in terms of synchrotron emission from a  
population of particles with a decreasing high-energy cutoff and  
moderate beaming (P01).  
                 
\noindent{\bf PKS~2201+044:} This BL Lac is hosted by a prominent  
elliptical galaxy at $z$=0.027. At one time, its showed an optical  
spectrum typical of a Seyfert 1, with broad (H$\alpha$, H$\beta$) and  
narrow ([OI], [OIII]) emission lines (Veron-Cetty \& Veron 1993). In  
the radio, it shows a core-jet morphology with total extension of  
5.6\arcmin\ (Ulvestad \& Johnson 1984; Laurent-Muehleisen et  
al. 1993). A bright and compact radio feature was found at 2.2\arcsec\  
from the nucleus. Subsequent \hst\ observations revealed an optical  
counterpart to the 2.2\arcsec\ hot spot and additional candidate knots  
in the inner parts of the jet (Scarpa et al. 1999). The bright optical  
2.2\arcsec\ knot is resolved by \hst\ in directions both parallel and  
perpendicular to the jet axis, with a complex structure.

\subsection{X-ray}   
  
\chandra\ observed \3c\ and \pk\ on 2002 August 8 and 2002 April  
27, respectively. The live times were 40~ks and 38~ks, yielding 37~ks
and 34.6~ks of useful exposure time after removing flaring background
events that occured during the observations. The background light
curves were extracted from source-free regions on the same chip of the
source. The data were analyzed using standard screening criteria. In
particular we used the version 3.3 of the
\verb+CIAO+ software package and the latest calibration files provided  
by the \chandra\ X-ray Center (\verb+CALDB v.3.2.1+).  
  
Both sources were observed at the aim point of the S3 chip of the  
ACIS-S detector. Since the sources were expected to have a bright  
core, the 1/8 subarray mode (frame time of 0.44 s) was used to reduce  
the effect of the nuclear pileup. In a circular region of radius  
1.5\arcsec\ the counts rates of the cores of \3c\ and \pk\ are 0.54 and  
0.29 c/s, respectively. According to \verb+PIMMS+, the estimated  
pileup percentages are relatively small, 8\% and 4\%, for the two  
sources.  The two X-ray images show a readout streak on the S3 chip  
that has been removed by the \verb+destreak+ tool, for imaging  
purpose.  
  
The cores fluxes and spectra were extracted using \verb+panda+ regions  
centered on the pixel with the highest counts. These regions, that  
resemble a pie with a slice removed, have an outer radius of 4\arcsec\  
and aperture angles of 260\deg\ and 270\deg\ for \3c\ and \pk,  
respectively.  The aperture angles were chosen to avoid the jet  
contamination.  
  
For the detected knots, fluxes and spectra were extracted from  
ellipses with axis lengths of 0.41\arcsec $\times$ 1.06\arcsec\ for  
\3c, and 0.43\arcsec $\times$ 0.85\arcsec\ for \pk, respectively,   
centered on the radio positions and with the major axis oriented  
perpendicular to the jet. The corresponding aperture corrections are  
2.7 for \3c, and 3.3 for \pk. The sizes of the ellipses represent a  
compromise between the need to maximize the number of X-ray photons  
while at the same time covering an area similar to the regions used at  
radio and optical, where the angular resolution is better. For the  
innermost radio/optical knots $\alpha$, where the contribution of the  
ACIS PSF's wings is likely to be strong, we used the same  
procedure. In all cases the background was evaluated in  
\verb+panda+ regions centered on the nucleus and with inner and outer  
radii such as to match the width of the source extraction region. The
choice of this background region ensures that the contribution of the
diffuse thermal emission and/or the wings of the PSF as a function of
azimuth are averaged. Figures~\ref{3c371in}--\ref{2201out} show the
ACIS-S images of the inner and outer jets.
  
The net count rates of the knots are reported in Table~1. We use   
the same nomenclature as specified in Scarpa et al. (1999) for the  
bright knots A and B in \3c\ and A in \pk, and the greek alphabet   
to identify the not previously named inner knots.  For knots  
$\alpha$, the closest to the core and only weakly detected, we only  
list conservatively a $3\sigma$ upper limit to the count rates. Knots  
$\beta$ at 1.3\arcsec\ and 1.6\arcsec\ from the cores were detected at  
$3\sigma$ over the background; for these, we also extracted ACIS  
spectra.    
  
The ACIS spectra of the cores and the jet knots have been analyzed  
within \verb+XSPEC v.11.3.2+. The spectra of components with a large  
number of counts (e.g., 100 or more) were grouped so that each new  
energy bin had at least 20 counts to enable the use of the $\chi^{2}$  
statistics.  Errors quoted throughout are 90\% for one parameter of  
interest ($\Delta\chi^2$=2.7). The unbinned X-ray spectra with $<$100  
counts (i.e., knot A for \3c\ and knots $\beta$ and A for \pk) were  
fitted using the C-statistic.  The uncertainties on the X-ray counts,  
$\sigma_N$, were calculated according to the formula  
$\sigma_N=[(\sigma_S)^2+(\sigma_B)^2]^{1/2}$, where $\sigma_S$ and  
$\sigma_B$ are the uncertainties of the source and the re-scaled by  
area background, respectively. In the regime of low counts, to  
evaluate the uncertainties we adopted the formula in Gehrels (1986):  
$\sigma = 1+(S+0.75)^{1/2}$.  
  
The X-ray fluxes for the individual knots are listed in Table~4. All  
fluxes are background-subtracted and corrected for finite aperture  
effects. They were also corrected for the expected contamination due
to the PSF wings of nearby knots. The total correction factors are
given in the Table. 
  
\subsection{\hst}  
  
The sources were observed by us with the WFPC2 and STIS on \hst\ in  
summer 2001. We also searched the \hst\ archive and found additional  
data. \3c\ has been the target of multiple observations related to 7  
different proposals, from 1994 to 2005, including, besides our  
observations, images with Near-Infrared Camera and Multi-Object  
Spectrometer (NICMOS) and the Advanced Camera for Surveys (ACS), the  
latter published by Perlman et al. (2006).  On the contrary, \pk\ was  
observed only with STIS and WFPC2 between 1998 and 2001. In many  
cases, the two sources were observed using different apertures and  
filters. The observation log of the \hst\ observations considered in  
this paper is presented in Table~2.   
  
We retrieved the data from the Multimission Archive at Space Telescope  
(MAST), reprocessed by the standard calibration pipeline. In particular,   
we downloaded the mosaic image of NICMOS, the flat-fielded images of ACS   
and STIS, and the c0f files of WFPC2. The subsequent reduction of the   
data was performed with the NOAO image processing software Image   
Reduction Analysis Facility (IRAF).  
  
In the case of multiple exposures, we stacked them together, removing  
the cosmic rays with the CRREJ task. In the inner part of the radio  
jets the optical emission is contaminated and/or overwhelmed by the  
light of the host galaxy. With the tasks ELLIPSE and BMODEL we fitted  
the host emission using a two-dimensional elliptical galaxy model,  
after masking the jet and diffraction spike areas.  The subtraction of  
the galaxy model from the stacked images revealed the complex  
morphology of the two jets also at optical wavelengths.   
  
Close to the nucleus, the galaxy subtraction is likely to leave  
imperfections. Indeed, in Figure~1 the feature in F555W alongside  
knots $\beta$ and B is most likely a residual of such an imperfect  
model, as it is not present in the other filters. In any case, it does  
not affect the measurement of the flux as the latter was extracted  
from a small region.   
  
We used squared apertures centered on the radio jet position to
extract the optical flux. Although the dimensions of these regions
change very slightly from one \hst\ detector to the other due to the
difference in pixel size, they are of the order of
$~\sim$0.3\arcsec. In the case of knots A, which are resolved both
longitudinally and transversally to the jet axis (see \S~3), we used
an extraction region with a total area similar to the X-ray extraction
region, or 0.3 arcsec$^2$. The background was estimated using several
identical apertures in adjacent regions to the extractions. Due to
background variation, we used the average and standard deviation of
the values estimated in the boxes.
  
Counts were converted to flux densities using the value in PHOTFLAM  
(the inverse sensitivity measurement) in the image headers. We  
corrected the extracted fluxes not only for the effects of the finite  
``aperture'' but also for the Galactic extinction. The corrections  
for the finite aperture were derived using the profile of the  
encircled energy of each detector. For the Galactic extinction we  
fitted the values reported in NED (Schlegel, Finkbeiner, \& Davis  
1998) with a power law and we extrapolated the value corresponding to  
the pivot wavelength of each filter.  In particular, the extinction  
corrections for \3c\ were about 2\% for NICMOS, 10\% for ACS,  
11\% for WFPC2 (F555W), and 46\% for STIS. For the observations of  
\pk\ we corrected the fluxes by 8\% for STIS, 9\% and 13\% for WFPC2  
(F702W and F555W, respectively). The optical flux densities in the
various cameras/filters are listed in Table~4.  The errors are of the
order of 3--30\% and, in the inner part of the jet, is likely
dominated by the uncertainties related to the galaxy subtraction.
  
For optical non-detections, we quote 3$\sigma$ upper
limits. Furthermore, the flat-fielded images of ACS are characterized
by a geometric distortion. For this reason we applied a
field-dependent correction: Using the plot of the Pixel Area Map (PAM)
reported at the \hst\ webpage www.stsci.edu/hst/acs/analysis/PAMS for
the WFC1 channel of the ACS, we found that the coefficient that has to
be applied for the observation of \3c\ is close to 0.985; to correct a
flux measured on the flat-fielded file it is necessary to multiply by
the PAM value and divide by the exposure time. 
  
In the NICMOS image, after subtracting the galaxy contribution, the  
jet axis of \3c\ appears to coincide with one of the diffraction  
spikes produced by the intense light from the core. The inner part of  
the jet is completely overwhelmed by the spike emission and no  
reliable estimate of the inner knot flux is possible. Although the  
spikes span the entire image and their intensities change irregularly  
with the distance from the central PSF, we note that 1) opposite  
spikes are approximately symmetric, and 2) their intensities decrease  
dramatically at $\sim$3\arcsec\ from the core.  That is, it is  
possible to roughly estimate the emission in the outer part of the jet  
after removing the contamination due to the spike. To perform this  
task, from the NICMOS image we subtracted the image itself rotated by  
180\deg.  This estimate has to be used with caution since spurious  
fluctuations can remain due to uncertainties in the relative intensity  
of the diffraction spikes.

\subsection{Radio}  
  
As part of our \chandra/\hst\ imaging program of these jets, we obtained  
new deep radio observations of the targets supplemented by a number of  
archival data sets. The data processing was done with AIPS (Bridle \&  
Greisen 1994) and DIFMAP (Shepherd et al. 1994) using standard  
procedures. Table~3 reports the log of the radio observations  
including archival data sets used for both sources.  
  
The new 22 GHz observations were obtained with the NRAO Very Large  
Array (\vla) as part of our observing program on large-scale X-ray jets  
(Cheung 2004; Sambruna et al. 2006).  Both objects were observed over  
a wide range of hour angles for uniform coverage of the ($u,v$) plane  
during separate observing runs -- \3c\ in the B-configuration and  
\pk\ with the hybrid AnB configuration to improve the  
north-south resolution for this lower declination source.  The total  
integrations of $\sim$3 hrs, and the off-source rms is near the  
theoretical noise limit of 1$\sigma$$\sim$0.1 mJy beam$^{-1}$.  
  
Comparably high resolution ($\sim$0.25'') radio maps at 1.4 GHz were  
obtained from \merlin\ (Table~3). A new full-track observation of  
\pk\ was obtained and for \3c, we use the full resolution  
version of the data originally presented in P01.  
  
To facilitate comparison with the largest scale structure (beyond knot  
A), we present 2.1\arcsec\ resolution maps of \3c\ at 5~GHz (from  
Wrobel \& Lind 1990). Radio emission beyond knot A in  
\pk\ is not quite visible in the existing data though lower  
resolution images show much extended structure (Ulvestad \& Johnston  
1984). We therefore obtained new subarcsecond-resolution \vla\ data at  
8.5 GHz for this object which details more of the extended radio tail.  
  
\noindent{\bf Polarization Images.}   
Additional polarization calibration of the 8.5 GHz \vla\ data for \pk\  
(Figure~\ref{polapk}) shows appreciable polarization structure in the  
jet. The jet is initially highly polarized ($\sim$15--20$\%$) with an  
inferred magnetic field transverse to the jet. At the jet flaring point  
of about 2\arcsec, there is more complex polarization behavior.  The  
observed polarization angle in the central portion of knot A is  
consistent with that observed at 5 GHz (Laurent-Muehleisen et al.  
1993). The agreement is within $\sim$10\deg\ which implies a small  
rotation measure of \simlt 76 rad m$^{-2}$ corresponding to a  
negligible polarization angle correction in our 8.5 GHz image.  The  
edges of the jet in these regions are very highly polarized. On the  
northern edge (fraction polarization $\sim$50$\%$), the inferred  
magnetic field orientation is parallel to the jet surface.  Toward the  
southern edge, the polarization is at the theoretical maximum for  
synchrotron radiation ($\sim$70$\%$) with no obvious alignment of the  
$B$-field vectors and the jet direction. At the spine of the jet in  
this region, the polarization is lower ($\sim$6--9$\%$), presumably  
from dilution because of the large beam size and cancellation from  
line-of-sight component of the magnetic field in the sheath layer.  
  
The polarization behavior in \3c\ (Figure~\ref{pola3c}) is similar to  
the case of \pk. Our analysis of an archival \vla\ 5 GHz data for \3c\  
shows similarly, highly polarized edges in knot A ($\sim$30--40$\%$)  
with inferred B-field vectors parallel to the jet surface.  We ignored  
the small correction +3\deg\ in the polarization angle at this  
frequency implied by the low rotation measure in this region of the jet  
(Wrobel \& Lind 1990).  The other datasets did not detect appreciable  
polarization in the jet.  In contrast to the \pk\ case, the B-field  
orientation in the \3c\ jet spine is parallel to the jet and about  
25$\%$ polarized. These data are consistent with the polarization map  
presented previously by O'Dea et al. (1988), and with the optical  
polarization image obtained with \hst\ ACS by Perlman et al. (2006).

\section{Imaging and Spatial Analysis}   
 
The 0.3--8~keV images of \3c\ and \pk\ are shown in
Figures~\ref{3c371in} and \ref{2201in}, and Figures~\ref{3c371out} and
\ref{2201out}, respectively. Also shown are the multiwavelength images of the
jets. Below we discuss separately the X-ray data for the cores and the
X-ray and multiwavelength observations of the inner and outer jets.

\subsection{The X-ray Cores} 

Inspection of the ACIS-S images suggests the presence
of diffuse X-ray emission around the cores of both sources. For a more
quantitative analysis we extracted the radial profiles of the core
region, using a series of concentric annuli centered on the core and
extending up to 400\arcsec\ on the ACIS CCD, with the jets
excised. For the extraction procedure, see Donato et al. (2004). The
jet and field point sources contributions were excluded from the
extraction regions.
  
The profiles are shown in Figure~\ref{core}. The core PSF fits well  
the inner parts (\ltsima8-10\arcsec) of the profiles, indicating the  
dominance of the core emission at such small radii (dashed  
line). Above 100\arcsec\ from the core, the profiles die into the ACIS  
background (dotted horizontal lines). However, excess   
flux over the PSF is present at $\sim$ 10\arcsec, most prominent in  
\3c. This feature suggests emission from a diffuse component, which we  
identify as ambient thermal gas. We thus fitted the excess flux with a  
$\beta$ model, which describes the thermal emission from gas in  
hydrostatic equilibrium (Cavaliere \& Fusco-Fermiano 1976).   
  
  
The results of the fits to the radial profiles with the composite PSF  
+ $\beta$ + background model are reported in Table~5; the model  
components and their sum (solid line) are plotted over the data in  
Figure~\ref{core}. The fitted core radii of the thermal component are  
5\arcsec, or $\sim$5 kpc, and 17\arcsec, or 9 kpc, for \3c\ and \pk,  
respectively, consistent with the X-ray haloes of FRIs.  The diffuse   
component is weakly detected in \pk\ (\ltsima$3\sigma$) and only by  
fixing the $\beta$ parameter at 0.67, typical of FRIs (Worrall \&  
Birkinshaw 1994), does the fit converge.   
  
We extracted the ACIS spectra of the diffuse emission in \3c, using   
\verb+panda+ extraction region between 10 and 30\arcsec. The background  
was chosen in a region on the same CCD free of serendipitae sources. A
good fit ($\chi^2_r=1.16/10$) of the background-subtracted spectrum is
obtained with a combination of a thermal component at soft energies
($kT=0.2^{+0.1}_{-0.2}$ keV, abundances fixed $Z=0.2/Z_{\odot}$) and a
power law at higher frequencies ($\Gamma=0.8\pm0.5$). The latter is
likely an instrumental feature: in fact, we detect a $\Gamma=0.8$ in
the \chandra, but not in the \xmm, spectra of diffuse emission around
other AGN (Lewis et al. 2007, in prep.). If true, it could represent
the onset of the non-thermal emission from relativstic particles in
the diffuse gas. The 0.3--8~keV observed flux of the diffuse thermal
component in \3c\ is F$_{0.3-8~keV}=5.1\times 10^{-14}$ \flux.

\subsection{The Jets} 
 
\begin{figure}[ht]  
\begin{center}
\hbox{
\includegraphics[width=18cm]{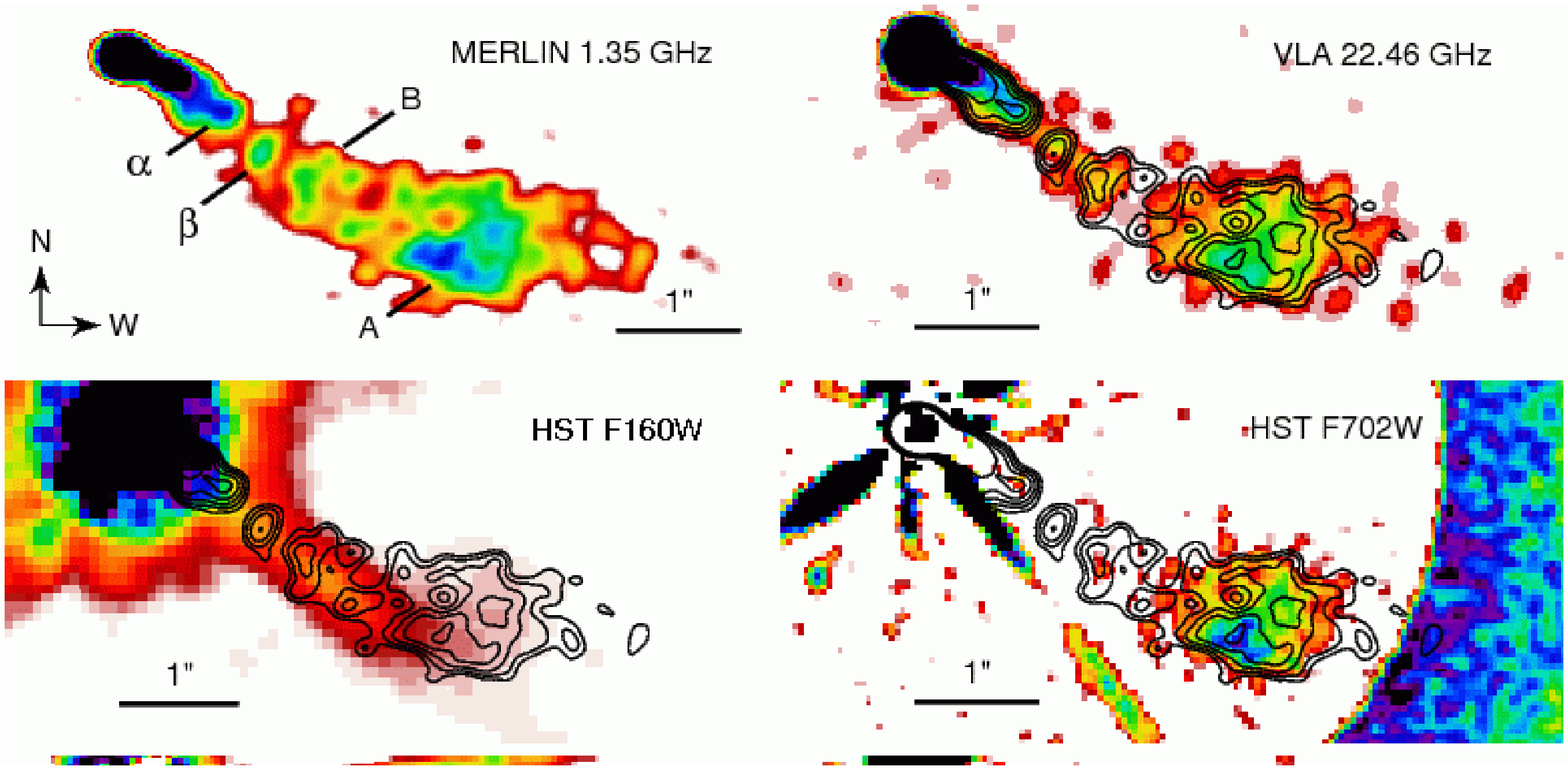}
}
\hbox{
\includegraphics[width=18cm]{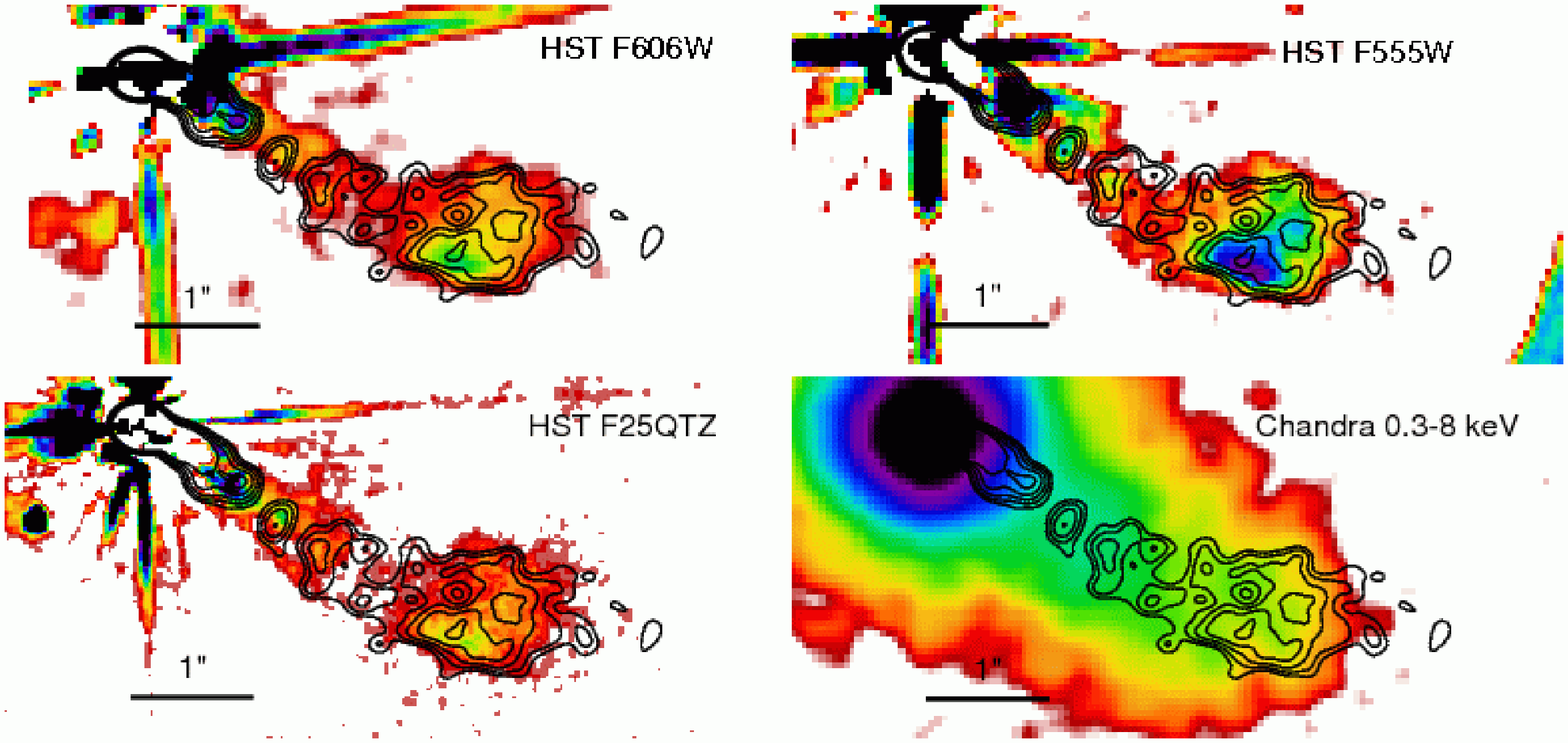}
}
\end{center} 
\caption{\footnotesize   
Images of the inner part of the jet of \3c\ at the various  
wavelengths. First row: \merlin\ (1.35 GHz) and \vla\ (22.46 GHz);  
Second row: NICMOS (F160W) and WFPC2 (F702W); Third row: ACS (F606W)  
and WFPC2 (F555W); Fourth row: STIS (F25QTZ- NUV-MAMA) and \chandra\  
0.3--8 keV.  The radio images have a resolution of $\sim  
0.18\times0.15$\arcsec\ and 0.2\arcsec, respectively.  The optical  
images have been convolved using {\it fgauss} in {\it FTOOLS} with  
elliptical Gaussian function of 1 pixel (2 pixels for STIS), with  
final resolution $\sim$ 0.15\arcsec. In the NICMOS image, the inner  
part of the jet is buried under a diffraction spike (see text). The  
X-ray image has been rebinned by a factor of 5 (final image pixel of  
0.1\arcsec) and then smoothed with the {\it csmooth} in {\it CIAO}  
with a circular Gaussian of 0.1\arcsec, with final resolution  
0.9\arcsec\ FWHM.  In all cases, the 1.35 GHz radio contours are  
overlaid on the color image. }  
\label{3c371in}  
\end{figure}

Figures~\ref{3c371in} and \ref{2201in} show a montage of the  
multiwavelength images of the inner parts (\ltsima 4\arcsec) of the  
\3c\ and \pk\ jets, respectively. The images are arranged in order of  
decreasing wavelength from radio to X-rays; the higher-resolution  
radio countours from \merlin\ are overlaid on each image.  
  
The outer parts of the jets on larger scales are shown in  
Figures~\ref{3c371out} and \ref{2201out}. The lower-resolution \vla\  
data are shown (left), and the countours overlaid on the \chandra\ map  
(right). The outer jet sections extend beyond the \hst\ fields, so no  
information is available from these data on their optical emission.  
  
Concentrating on the inner jets, Figures~\ref{3c371in} and  
\ref{2201in} show that the radio jets are detected at higher  
frequencies. For \3c\ we show the galaxy subtracted NICMOS image. As  
mentioned above, the inner parts of the jets are located below one of  
the PSF spikes; however, diffuse IR emission is clearly seen from the  
region around knot A.  
  
  
In both sources, the jets share a similar overall multiwavelength  
morphology: while the radio-to-optical/UV jets morphologies track each  
other well, the X-ray emission is concentrated near the core, and  
fades toward the end of the jet. This multiwavelength morphology is  
very similar to what observed in general for FRIs (Worrall et  
al. 2001), where the dominant emission mechanism is synchrotron.  
However, although the cores are quite bright at X-rays in both sources  
(see \S~4),  
the contamination of the core PSF wings to the X-ray emission from the
inner X-ray jets is negligible due to the choice of the background
extraction region.
  
To better quantify the jet properties we extracted profiles along the  
main jet axis (longitudinal profiles) and perpendicular to it  
(transverse profiles) at three wavelengths: radio (1.35~GHz for  
\3c\ and 1.41 GHz for \pk), optical (F555W), and X-rays (0.3--8~keV).   
For \3c, the optical images were convolved using {\it fgauss} in {\it
FTOOLS} with elliptical Gaussian functions of 1 pixel (2 pixels for
STIS), with final resolution $\sim$ 0.15\arcsec\ matched to the
radio. For \pk\ the radio resolution is 0.25\arcsec; thus, its \hst\
images were convolved with a Gaussian function of 2 pixels to achieve
matching resolution to the radio. For both sources, the ACIS images
were rebinned by a factor of 0.2, yielding an image pixel size of
0.1\arcsec, and smoothed with \verb+csmooth+ with a sigma of 1.5
pixels, with final resolution of $\sim$ 0.9\arcsec. At this
resolution, some of the closest knot pairs (e.g., $\beta$ and B in
\3c) will be blended. Another issue is that, while the FWHM of any
point source in the field is close to 1\arcsec, the sub-pixel imaging
artificially forces each X-ray event to be confined to a 0.1\arcsec
$\times$ 0.1\arcsec\ pixel and many of these will lie outside the
extraction region.  This increased Poisson noise could produce
spurious features, e.g., the double-peak structure in the longitudinal
X-ray profile of \pk\ around knot A (Figure~\ref{pklongi}).  
  
The longitudinal profiles are shown in Figures~\ref{3clongi} and
\ref{pklongi}. In the radio and optical, the profiles were extracted
by collapsing the flux of the jet onto a box 4\arcsec-long and
0.15\arcsec-wide for \3c\ and 0.25\arcsec\ for \pk.  In the X-rays we
used boxes 4\arcsec-long and 1\arcsec-wide.  The transverse profiles
(Figures~\ref{3ctras} and \ref{pktras}) were extracted from boxes
3.5\arcsec-long and 0.15\arcsec-wide for \3c\ and 0.25\arcsec-wide for
\pk\ for the optical and the radio. In the X-rays, boxes
0.5\arcsec-wide were instead used, representing a compromise between
matching the ACIS resolution and the need to separate the knot
emission as much as possible. The boxes were placed across the jet at
given distances from the core (see below and figure captions), and
collapsing the flux onto the main axis of the box.

\subsubsection{3C~371}
  
\begin{figure}[ht]  
\centerline{\includegraphics[height=2.4in]{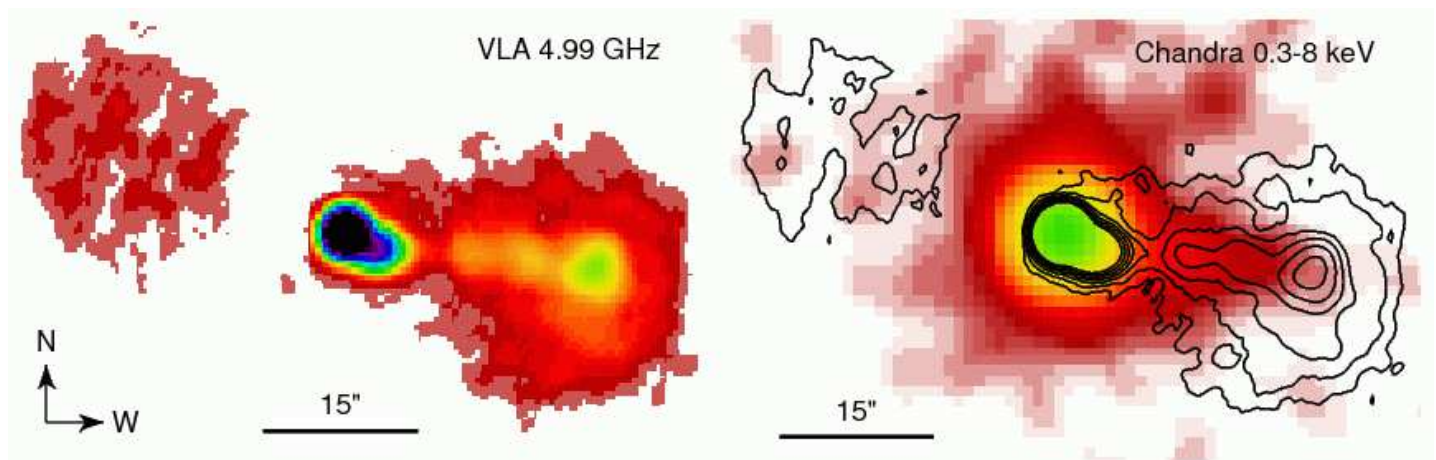}}  
\caption{\footnotesize   
Images of the outer part of the jet of \3c.  
Left: \vla\ (5~GHz); Right: \chandra\ 0.3--8 keV.  The radio image  
has a resolution of 2.1\arcsec. The X-ray image was rebinned by a  
factor of 2 (final image pixel of 1\arcsec) and smoothed  
with a circular Gaussian of 2\arcsec.}  
\label{3c371out}  
\end{figure}

\begin{figure}[ht]  
\begin{center}
\hbox{
\includegraphics[width=18cm]{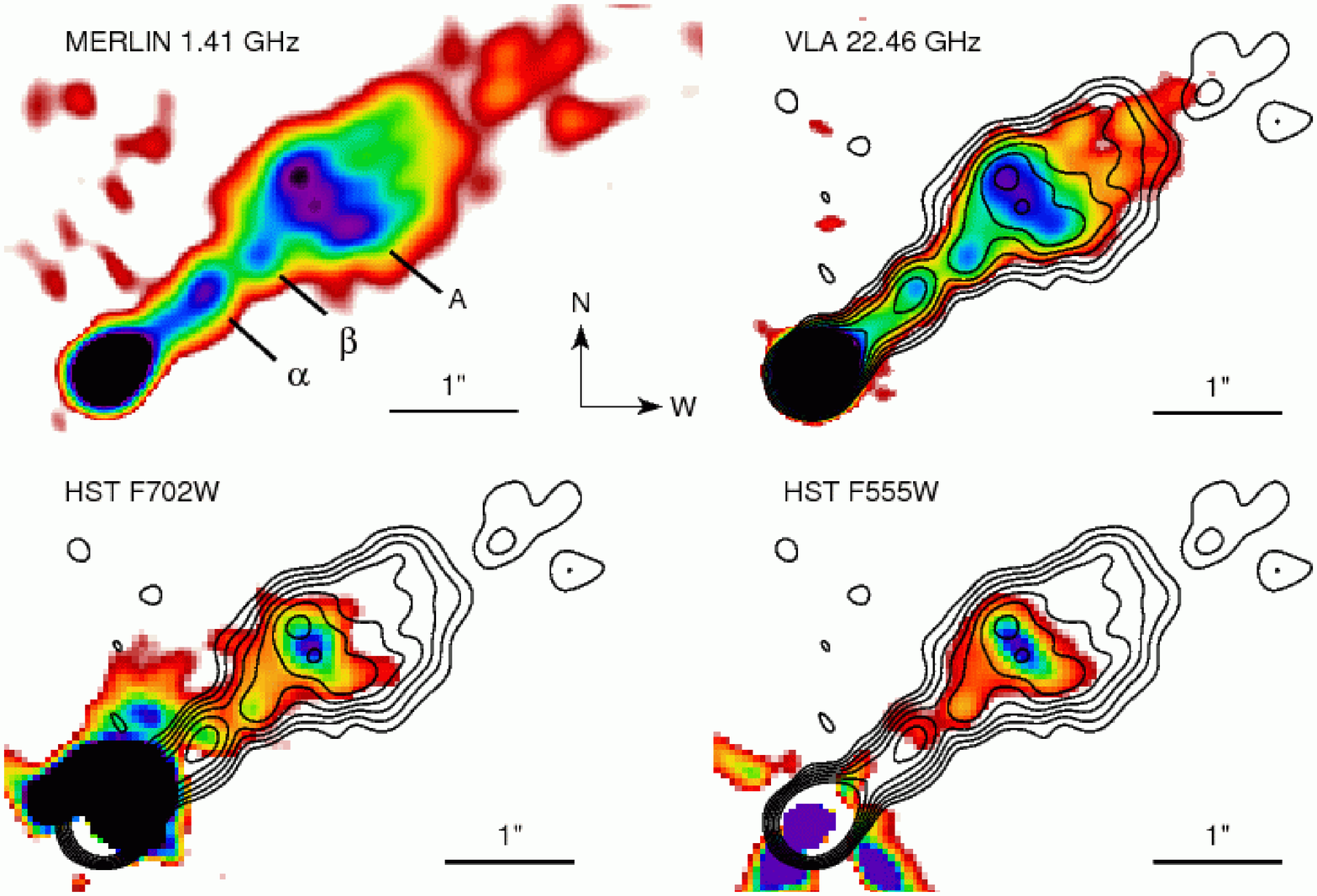}
}
\hbox{
\includegraphics[width=18cm]{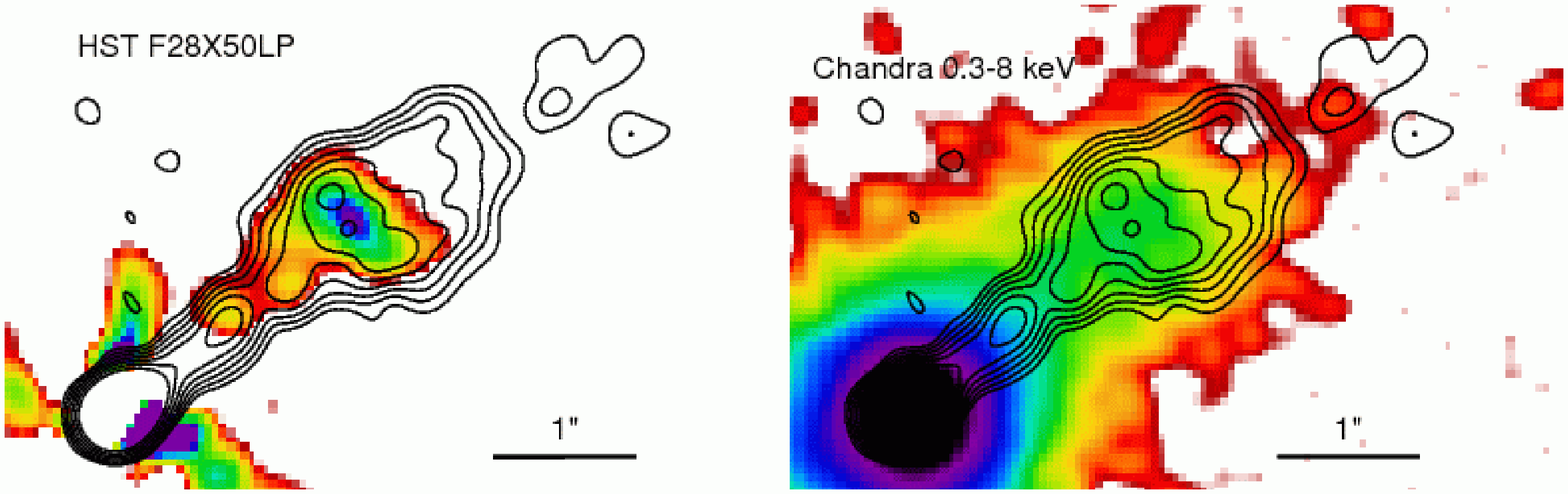}
}
\end{center} 
\caption{\footnotesize   
Images of the inner part of the jet of \pk.   
First row: \merlin\ (1.41 GHz) and \vla\ (22.46 GHz);    
Second row: WFPC2  (F702W and F555W);  
Third row: STIS (Broadband F28X50LP) and \chandra\ 0.3--8 keV.  
The radio images have a resolution of 0.25\arcsec. The optical and  
X-ray images have been obtained using the same procedure adopted for  
\3c\ (see Figure~\ref{3c371in}). The 1.41 GHz radio contours are overlaid on  
all color images.    
}  
\label{2201in}  
\end{figure}

\begin{figure}[ht]  
\centerline{\includegraphics[height=2.4in]{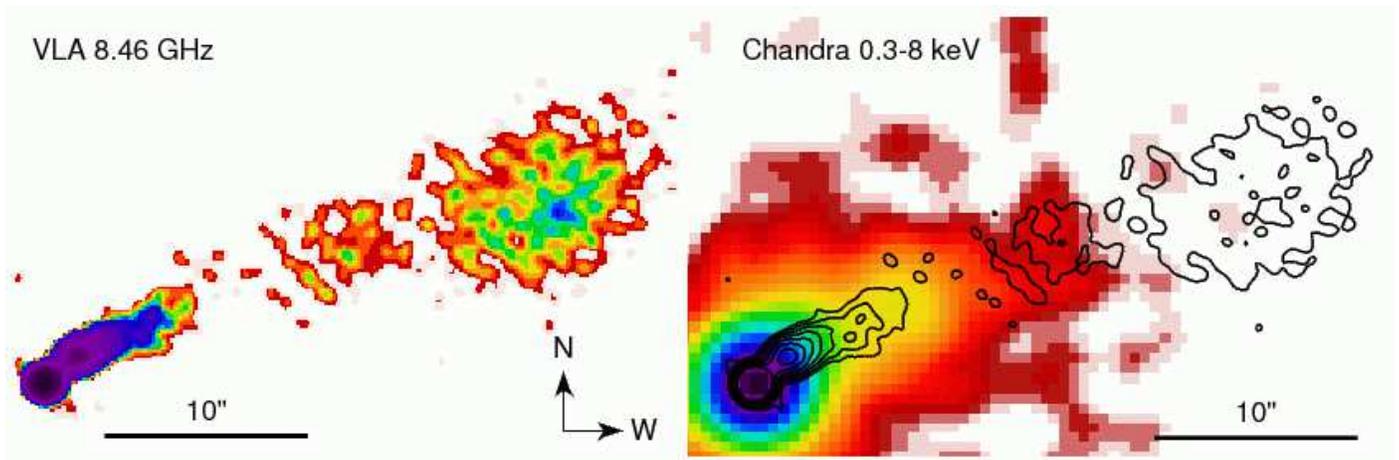}}  
\caption{\footnotesize   
Images of the outer part of the jet in \pk.  
Left: \vla\ (8.46 GHz); Right: \chandra\ 0.3--8 keV.  
The radio image  
has a resolution of 0.75\arcsec. The X-ray image was   
smoothed with a circular Gaussian of 1\arcsec.}  
\label{2201out}  
\end{figure}

\begin{figure}[ht]  
\begin{center}  
\hbox{  
\includegraphics[width=3.5in]{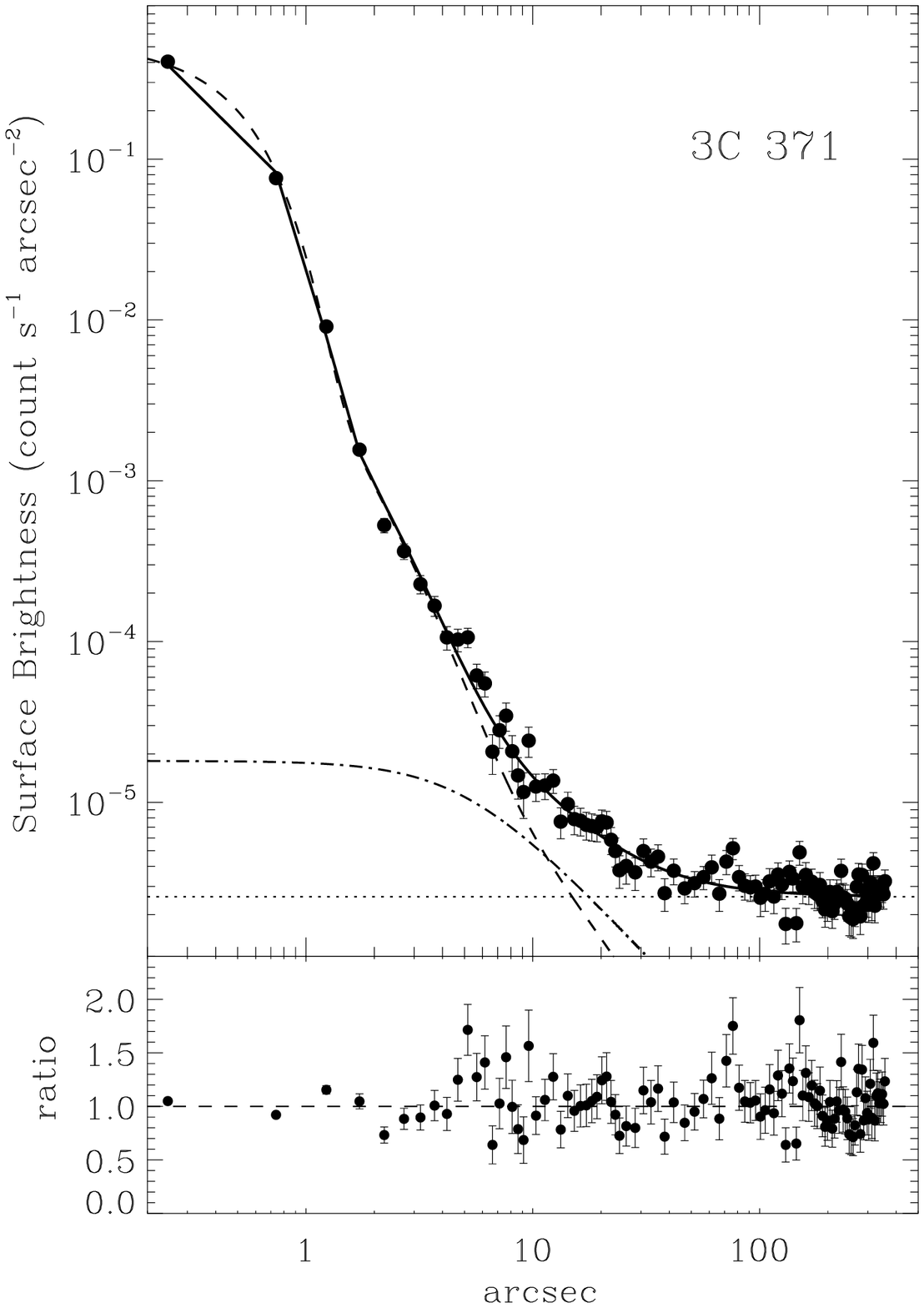}\includegraphics[width=3.5in]{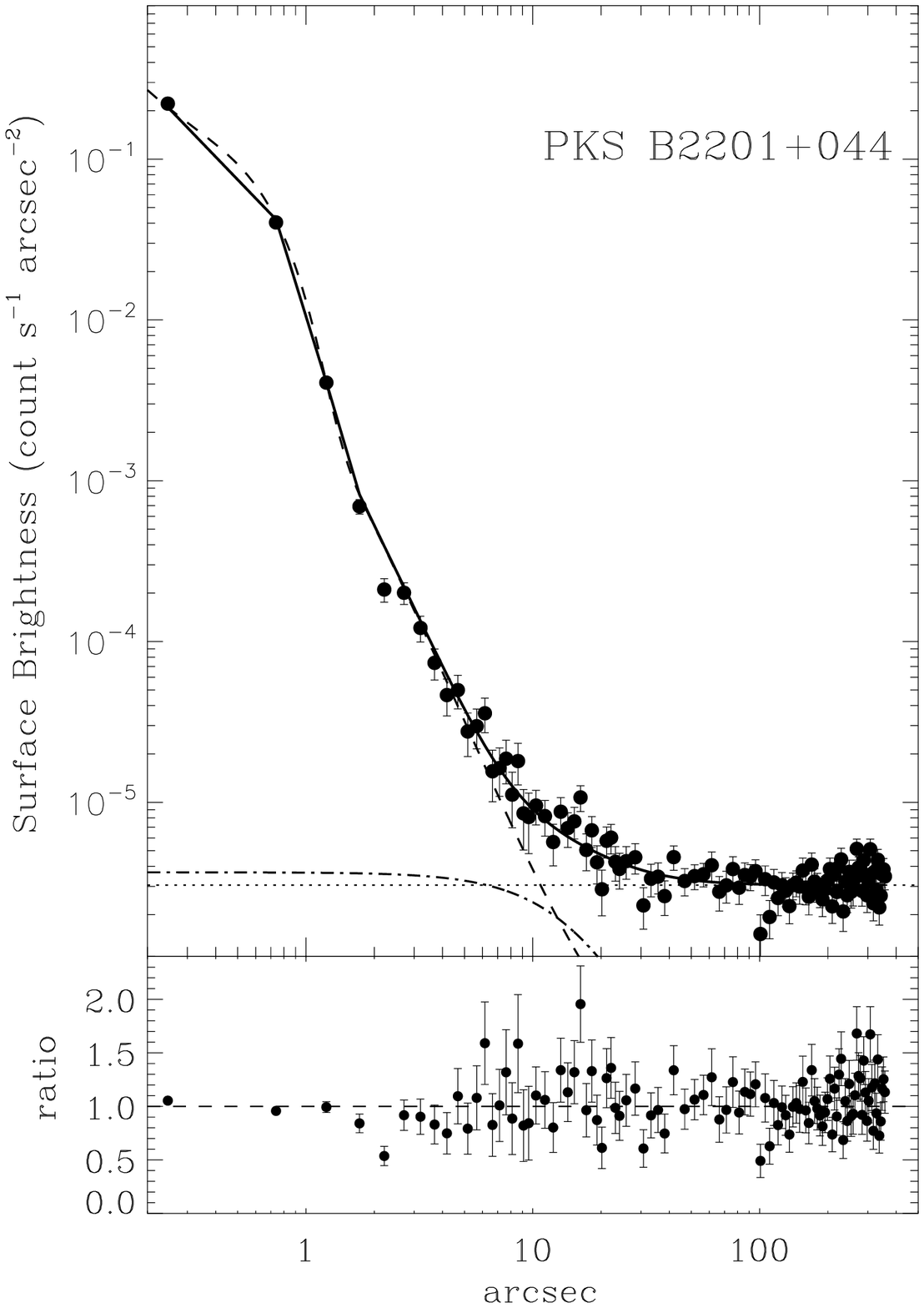}  
}  
\end{center}  
\caption{\footnotesize  
{X-ray (0.3-8 keV) radial profiles of the two sources. In the top   
panels, the dashed line represents the instrumental PSF, and  
the dot-dashed line the $\beta$-model describing the diffuse  
emission. The continuous thick line is the total model, while the thin  
dotted horizontal line is the background. The residuals of the  
best-fit model are shown in the bottom panels.   
}}  
\label{core}  
\end{figure}

\begin{figure}[ht]  
\begin{center}  
\hbox{  
\centerline{\includegraphics[width=11cm]{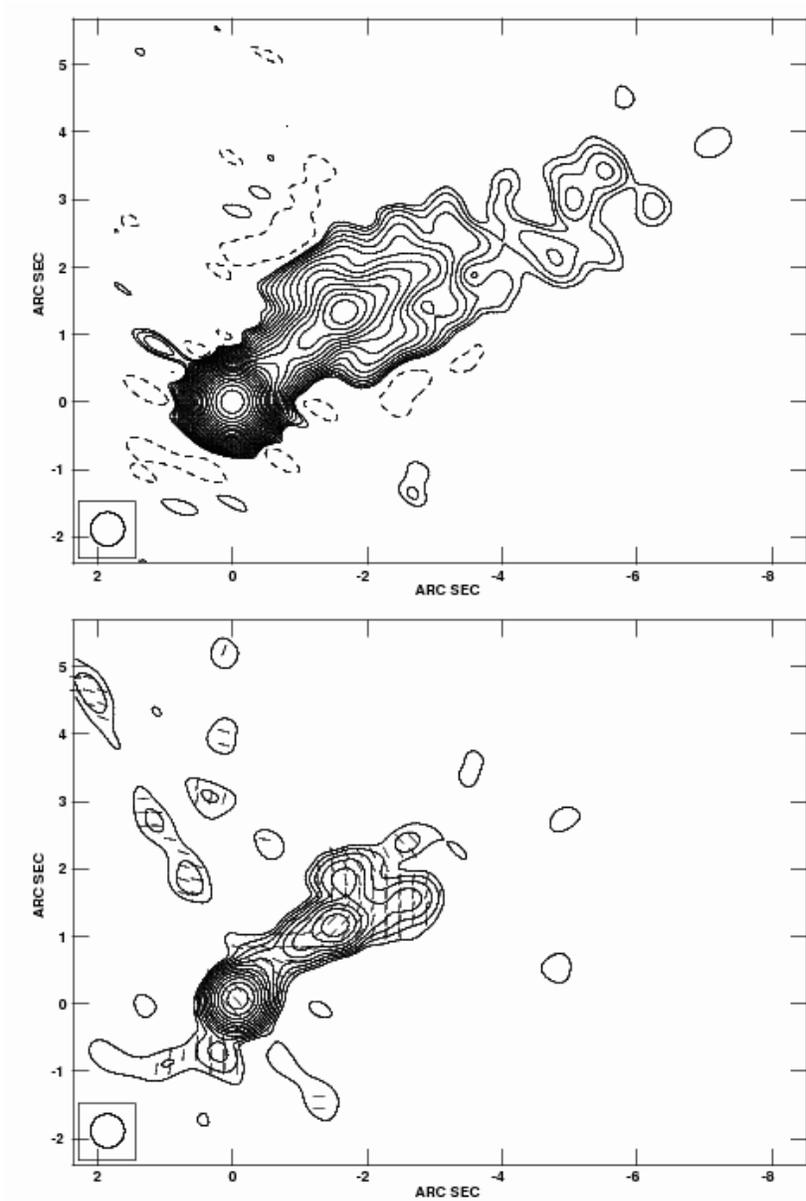}}  
}  
  
\end{center}  
\vspace{-1.0cm}  
\caption{\footnotesize  
{\vla\ total intensity [I; top] and polarization [P; bottom] images of
PKS~2201+044 at 8.5 GHz. The short solid lines in the bottom panel
represent the observed electric field. Contour levels begin at 0.11
(I) and 0.125 (P) mJy/bm and increase by factors of $\sqrt{2}$ to
peaks of 306.5 (I) and 7.2 (P) mJy/bm. The images are
``super-resolved'' with a 0.5\arcsec\ beam which is $\sim$1/2 of the
naturally weighted beam.  }}
\label{polapk}  
\end{figure}

\begin{figure}[ht]  
\begin{center}  
\hbox{  
\centerline{\includegraphics[width=4.5in]{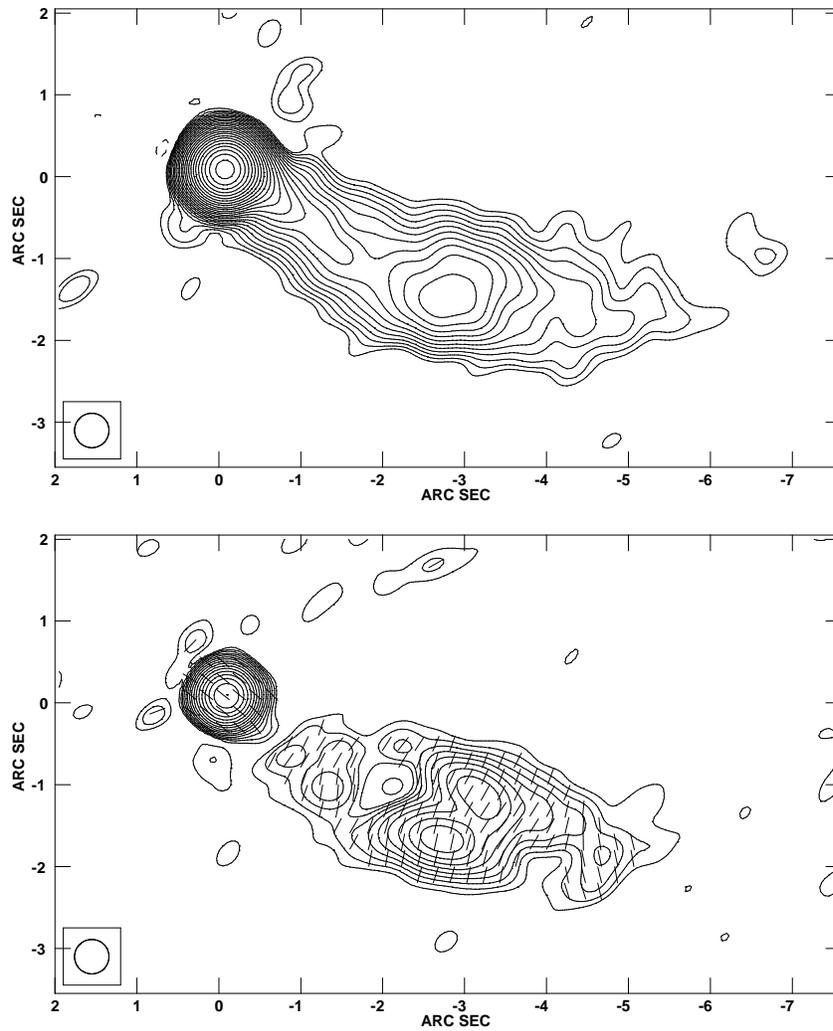}}  
}

\end{center}  
\vspace{-1.0cm}  
\caption{\footnotesize  
{\vla\ total intensity [I; top] and polarization [P; bottom] images of
3C371 at 5 GHz. The short solid lines in the bottom panel represent
the observed electric field. The contour levels begin at 0.45 (I) and
0.25 (P) mJy/bm and increase by factors of $\sqrt{2}$ to peaks of
1577.5 (I) and 45.3 (P) mJy/bm. The restoring beam of 0.42 arcsec is
plotted on the bottom left of each panel. The tick marks indicate the
orientation of the observed electric vector position angles.  }}
\label{pola3c}  
\end{figure}

\begin{figure}[ht]  
\begin{center}  
\hbox{  
\noindent{\includegraphics[height=16cm]{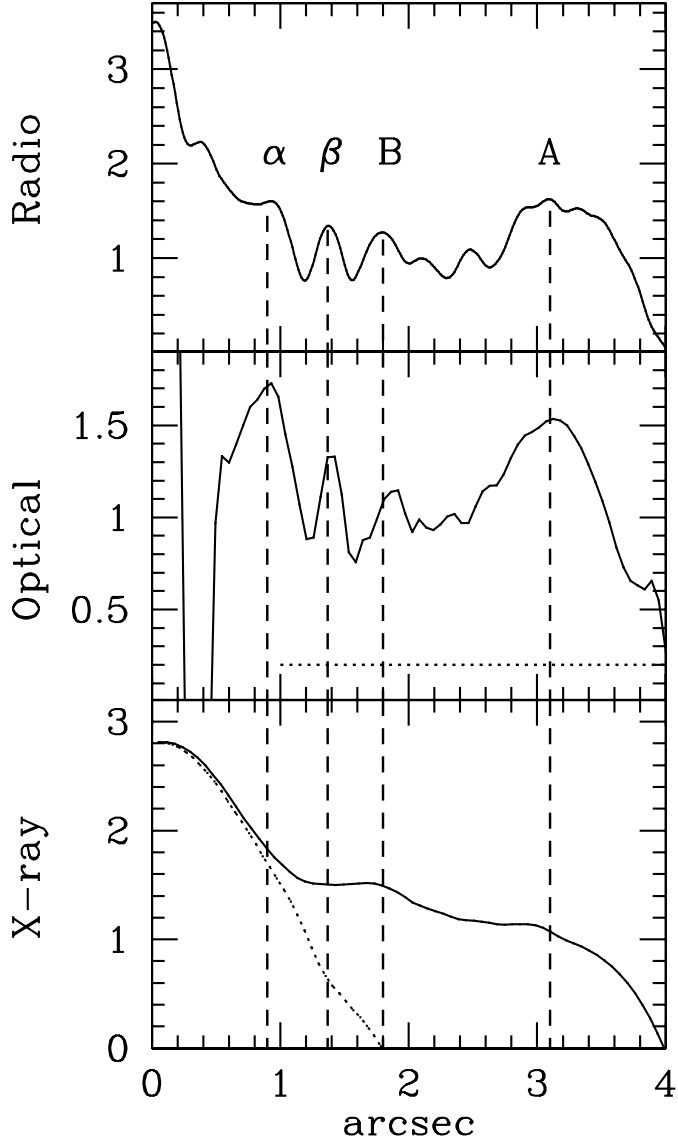}}  
}  
  
\vspace{0.1cm}  
  
\end{center}  
\vspace{-1.0cm}  
\caption{\footnotesize  
{Longitudinal profiles at three wavelengths for the jet of  
\3c: Radio at 1.35 GHz, optical at 5443~\AA, and X-ray in the 0.3-8 keV  
range (see \S~3.2 for the technical description of the images). The  
units of the y-axis, in logaritmic scale, are Jy/beam at radio, and  
counts/pixel at optical and X-ray. The dashed line in the central  
panel marks the optical background. The radio and X-ray backgrounds  
(including the thermal emission in the latter) are off scale, more  
than one order of magnitude lower than the jet emission. Also shown in  
the X-ray profile is the profile at an azimuth of 180\deg\ from the  
jet. The lack of data around 0.4\arcsec\ in the optical profile is an  
artifact due to the galaxy subtraction.  The uncertainties on the  
fluxes are of the order of 5--30\% for the X-rays, 3--30\% for the
optical, and 12--15\% for the radio. 
}}  
\label{3clongi}  
\end{figure}

\begin{figure}[ht]  
\begin{center}  
\hbox{  
\noindent{\includegraphics[height=16cm]{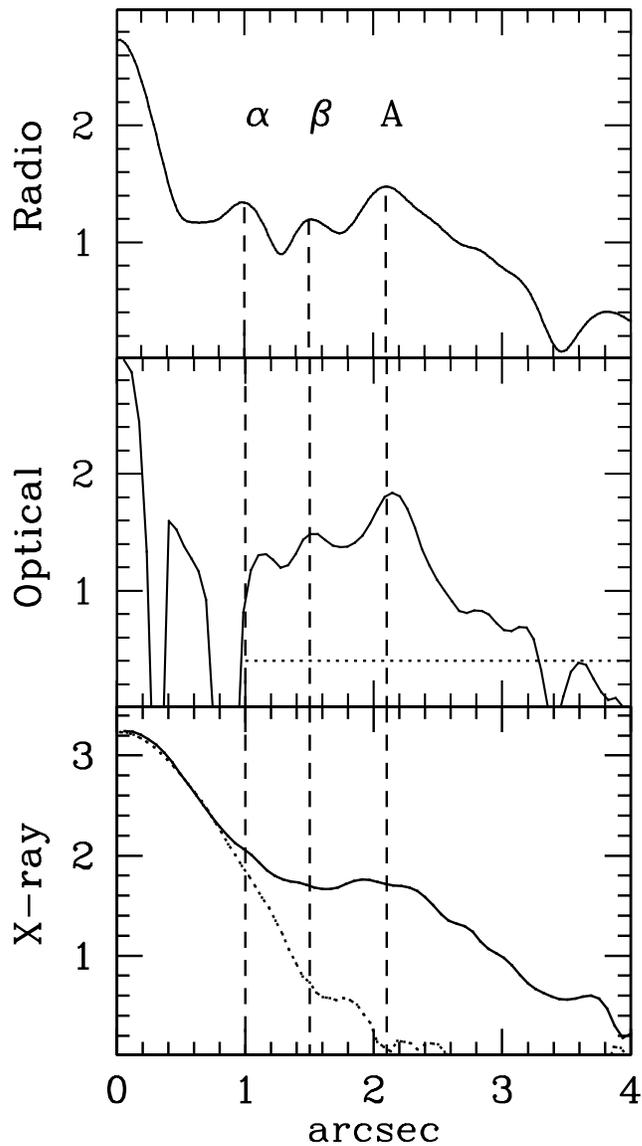}}  
}  
  
\end{center}  
\vspace{-1.0cm}  
\caption{\footnotesize  
{Longitudinal profiles at three wavelengths for the jet of  
\pk: Radio at 1.41 GHz, optical at 5443~\AA, and X-ray in the 0.3-8 keV  
range (see comments of Figure~\ref{3clongi}). Also shown in the X-ray
profile is the profile at an azimuth of 180\deg\ from the jet. The
lack of data around 0.3\arcsec\ and 0.9\arcsec\ in the optical profile
is an artifact due to the galaxy subtraction. Uncertainties as for
Figure~\ref{3clongi}. }}

\label{pklongi}  
\end{figure}

\begin{figure}[ht]  
\hbox{  
\includegraphics[bb=100 160 375 705,clip=,height=13cm]{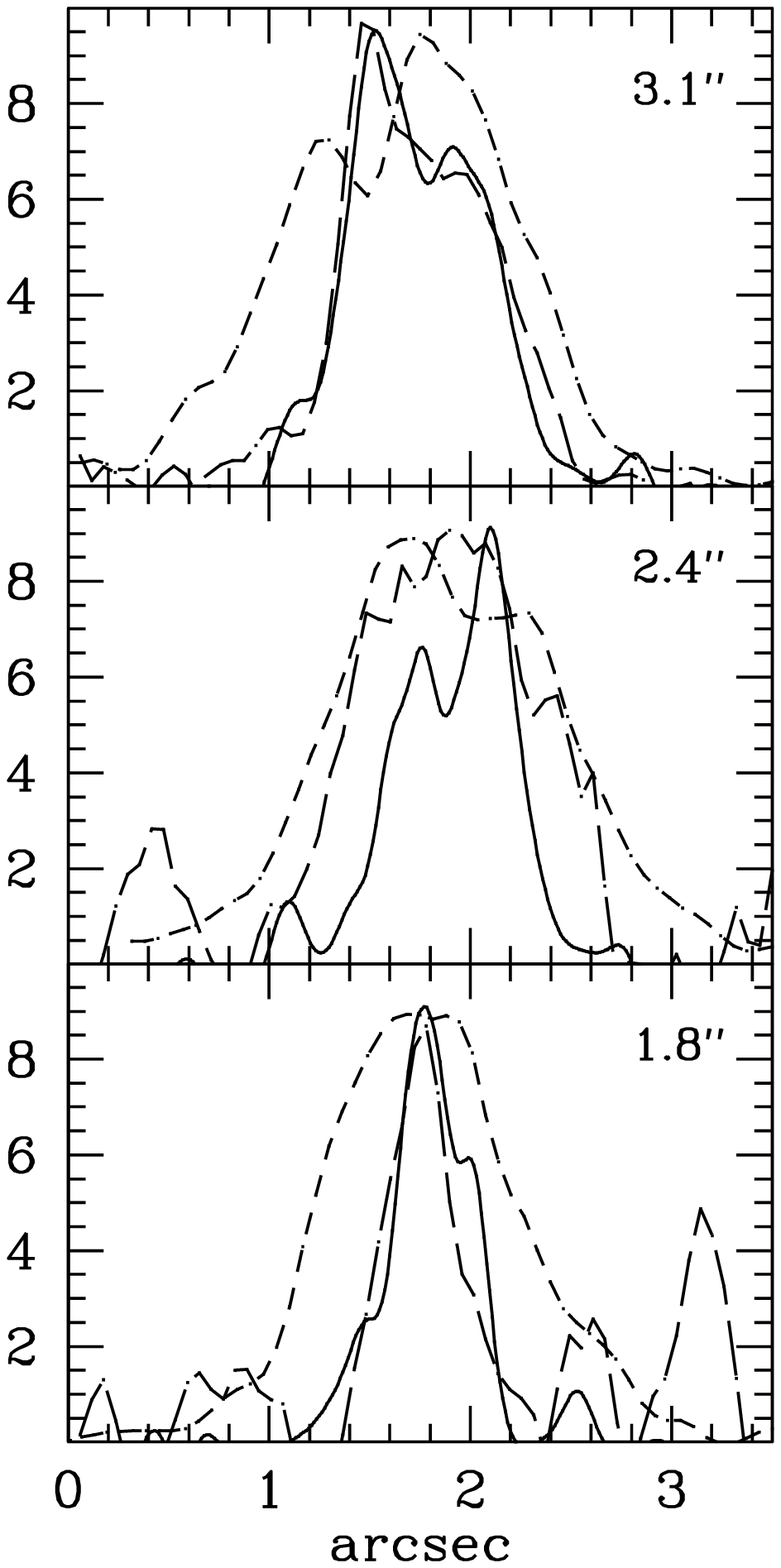}  
\includegraphics[height=6cm]{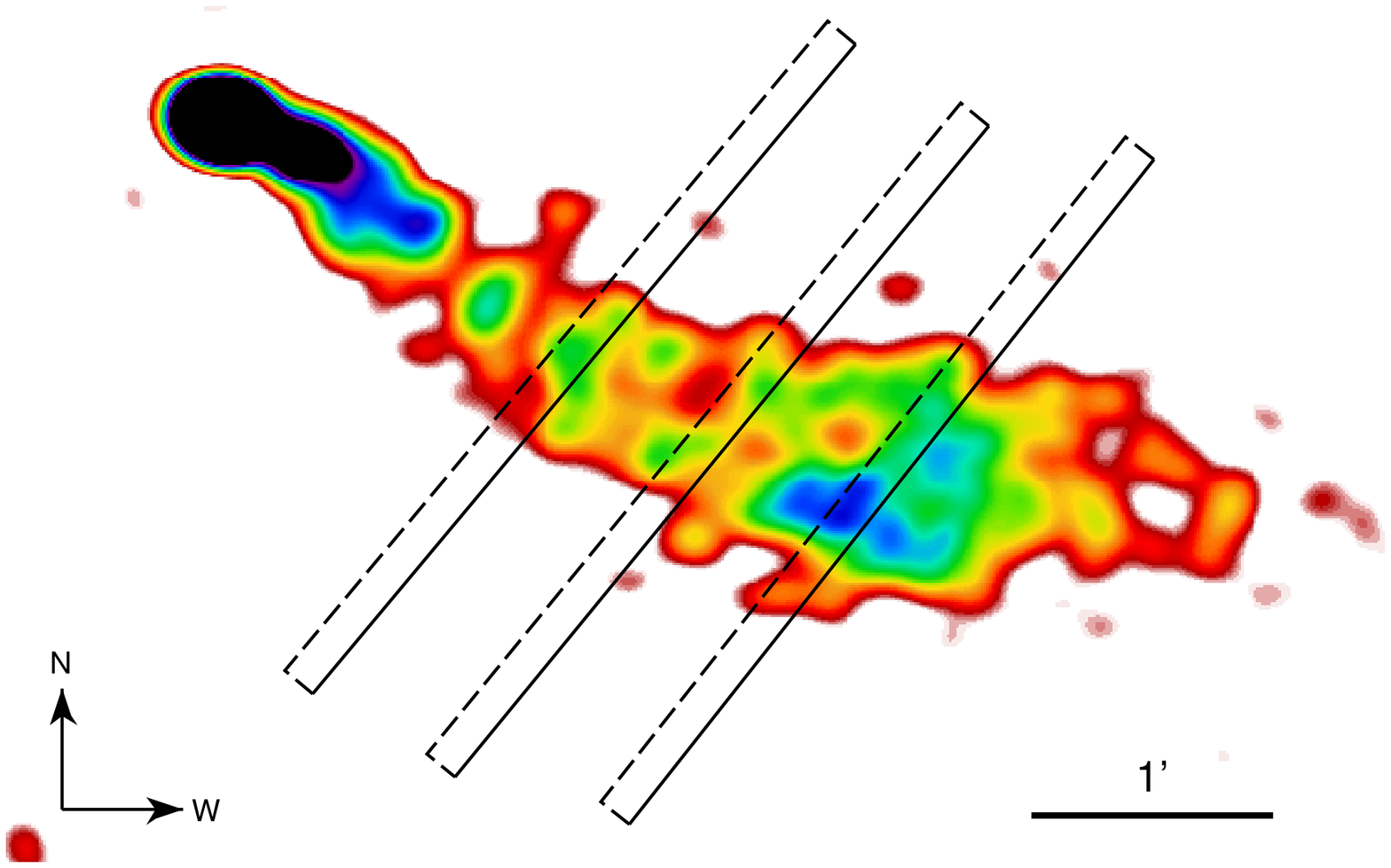}  
}  
%
%
\caption{\footnotesize  
{Normalized transverse profiles at three wavelengths for the jet of
\3c\ at a distance 3.1\arcsec\ (knot A, top), 2.4\arcsec\ (center),
and 1.8\arcsec\ (knot B, bottom). In all panels, the y-axis is
expressed in arbitrary units, the solid line is the radio, the
long-dashed line the optical, and the short-dashed line the X-rays.
The profiles are extrated from the same images used for the profiles
in Figure~\ref{3clongi}.  Also shown in the companion panel are the
positions of the extraction regions on the radio image. Uncertainties
as for Figure~\ref{3clongi}.  }}
\label{3ctras}  
\end{figure}

\begin{figure}[ht]  
\begin{center}  
\hbox{  
\includegraphics[bb=100 160 375 705,clip=,height=13cm]{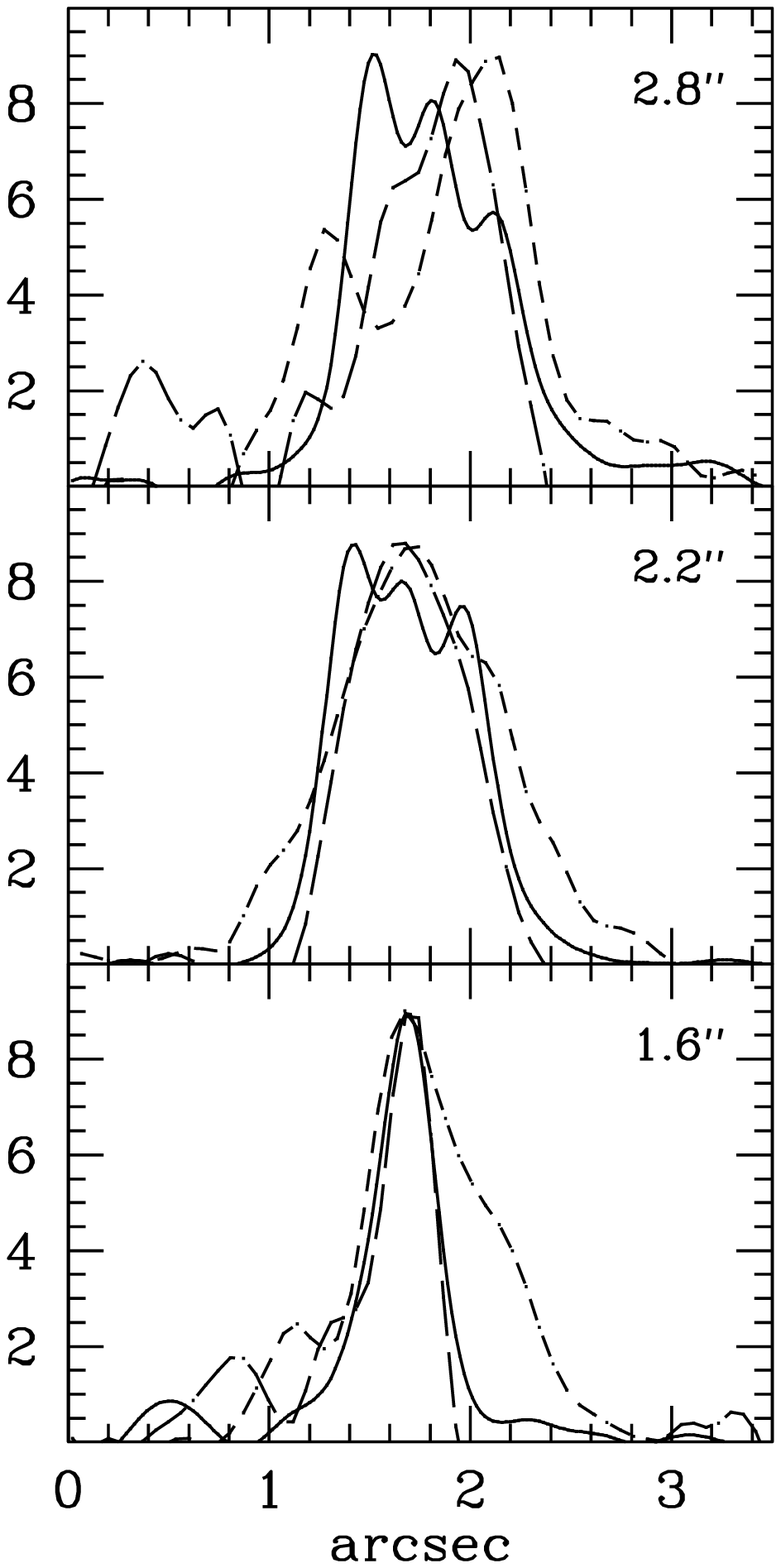}\includegraphics[height=6cm]{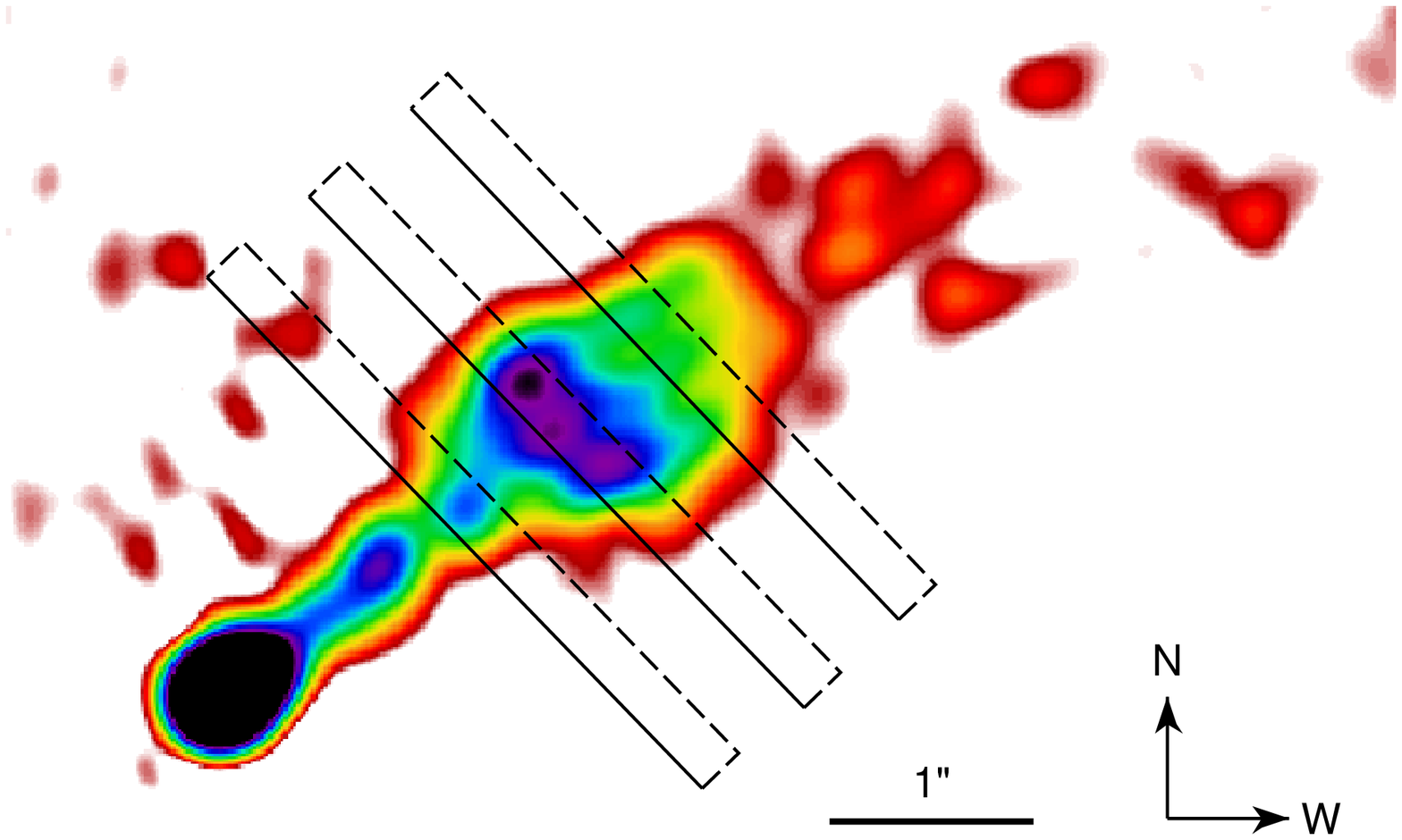}  
}  
%
%
\end{center}  
\vspace{-1.0cm}  
\caption{\footnotesize  
{Normalized transverse profiles at three wavelengths for the jet of
\pk\ at a distance 2.8\arcsec\ (top), 2.2\arcsec\ (knot A, center),
and 1.6\arcsec\ (knot $\beta$, bottom). In all panels, the y-axis is
expressed in arbitrary units, the solid line is the radio, the
long-dashed line the optical, and the short-dashed line the
X-rays. The profiles are extrated from the same images used for the
profiles in Figure~\ref{pklongi}. Also shown in the companion panel
are the positions of the extraction regions on the radio
image. Uncertainties as for Figure~\ref{3clongi}. }}
\label{pktras}  
\end{figure}

The inner jet is shown in Figure~\ref{3c371in} at the various  
wavelengths. We confirm our previous result (P01) that the  
radio/optical jet of \3c\ emits at X-rays in the deeper \chandra\  
image. In addition, we find faint X-ray emission from the outer jet,  
beyond feature A at \gtsima 4\arcsec\ (Figure~\ref{3c371out}).  

From radio to IR/optical and UV, the jet emission appears concentrated
in a few bright knots at $\sim$ 0.9\arcsec\ (knot $\alpha$) and
1.3\arcsec\ (knot $\beta$), then fades somewhat around knot B at
1.8\arcsec, and picks up again in the lobe near knot~A. The X-ray
emission has a broad shoulder in correspondence to knots $\beta$ and
B, beyond which it decreases.
  
At the higher angular resolution of \merlin\ and \hst, ``knot'' A is
resolved into a more complex structure. Here, the emission is spread
over a $\sim$ 1\arcsec-long feature piled against the southeast
intensity contours of the radio lobe. The radio and optical emission
then declines gradually from the southern to the northern border of
the emission region (knot A).  The poorer \chandra\ resolution
prevents a detailed mapping of the faint X-ray
emission. Interestingly, the jet changes direction at this location
(Figure~\ref{3c371out}). This strongly suggests that knot A is the
site of an impact of the jet against some external dense medium, where
shocked plasma emits optical and X-rays as a result of the compressed
magnetic field and accelerated electrons.
  
The longitudinal profiles of the inner jet are shown in  
Figure~\ref{3clongi}. The extraction box was located at a position  
angle PA=55\deg; the four main radio/optical features are labelled  
(see Table~1). This Figure emphasizes the overall good correspondence  
between the radio and the optical, which peaks in correspondence of  
knot $\alpha$ and structure A; again, the X-ray profile fades  
downstream of knot B.  
  
Emission profiles transverse to the jet axis taken at the position of
knot B, structure A, and at an intermediate location (2.4\arcsec)
between the two are shown in Figure~\ref{3ctras}. The main apparent
feature is that the profiles are resolved at all three wavelengths. If
the three X-ray profiles are approximated by Gaussians, the ensuing
FWHMs are 1.0--1.3\arcsec. After deconvolving the 0.9\arcsec\ of the
image PSF, the intrinsic region size for the resolved knot is
0.6--0.9\arcsec.
  
\begin{figure}[ht]  
\begin{center}  
\hbox{  
\centerline{\includegraphics[width=18cm]{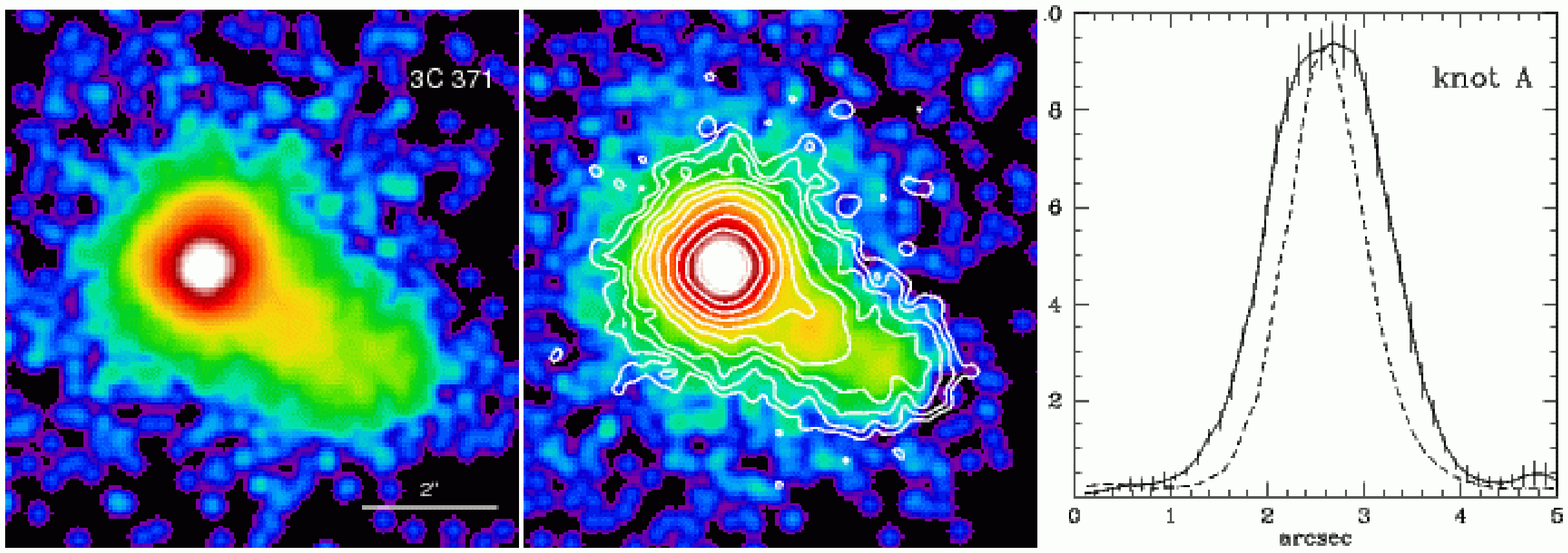}
}}

\vspace{0.5cm}  
\hbox{  
\centerline{\includegraphics[width=18cm]{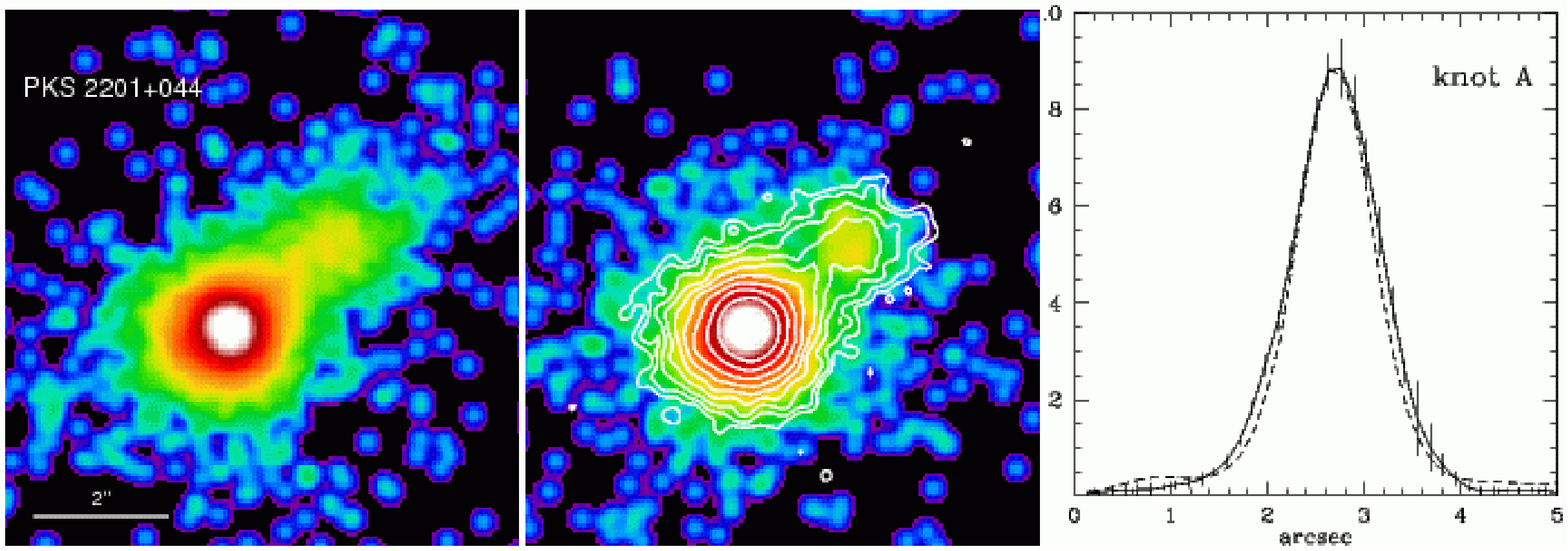}
}}  
  
\end{center}  
\vspace{-1.0cm}  
\caption{\footnotesize  
{Observed (left) and simulated (middle) ACIS-S images of the jets of  
\3c\ (top) and \pk\ (bottom); see text for more details. The contours  
of the observed images are overlaid on top the simulated ones to show  
the presence of non-zero X-ray emission from the ``edges'' of the jets  
(see text). All images have been rebinned by a factor of 10 (final  
image pixel of 0.05\arcsec) and then smoothed with the {\it  
csmooth} in {\it CIAO} with a circular Gaussian of 0.1\arcsec. The  
right panels show the radial profiles of knots A from the observed  
(continuous curve with errors) and simulated (dashed) images.}}  
\label{simuls}  
\end{figure}

This extension is not an effect of the ACIS PSF. To illustrate this,
using \verb+CHART+ and \verb+MARX+ we simulated the 37~ks ACIS
observation of \3c\ including the core and extended thermal emission
in addition to the jet.  The core and knots A and B were assumed to be
point-like, while the thermal emission around the core was described
by a beta-model (Cavaliere \& Fusco-Fermiano 1976) with the observed
characteristic radius of 5\arcsec\ and the $\beta$ parameter fixed at
0.5 (see \S~3.1).  The observed ACIS-S spectra were used to model the
various components (i.e., core and knots). Figure~\ref{simuls}, top
panels, shows the observed ACIS image (left) and the simulation
overlaid with the observed X-ray contours (center). Note the larger
X-ray contours compared to the simulated image; the excess flux of the
observation with respect to the the simulation is significant at
P$>$99.99\%. The transverse profiles of the observed and simulated
knot A are compared in the top-right panel of
Figure~\ref{simuls}. Thus, we conclude that the large aspect ratio of
the jet at X-rays is real and significant.
 
The only X-ray jet resolved so far orthogonally to its axis is in the
close-by FRI Centaurus~A, in a long (95~ks) \chandra\ ACIS-S exposure
(Kataoka et al. 2006); it has a flat-topped profile suggestive of
limb-brightening, previously observed at radio frequencies only
(Giovannini et al. 2000). The opening angle of the jet in X-rays is
half as large as at radio (Kataoka et al. 2006).

As apparent from Figure~7, the polarization images show the presence
of a structured jet. There are two peaks in the linear polarization
map separated by $\sim$ 1\arcsec\ in correspondence to knot A.
Moreover, the magnetic field lines are denser toward the edges of the
jet, indicating a ``spine/wall'' jet structure, which is also observed
with \hst\ (Perlman et al. 2006). This raises the possibility that the
X-ray spectrum may be different at the jet center and periphery.  We
will investigate this issue in \S~4.2.1 referring to the central and
lateral parts of the jet as ``spine'' and ``edges''; see
Figure~\ref{edges} and \S~4.2.1 for their definition.

  
We turn now to the outermost part of the jet at \gtsima  
4\arcsec. Figure~\ref{3c371out} shows how the radio jet changes  
direction after structure A, and has three knots of emission  
terminating in a diffuse lobe at 25\arcsec\ from the  
core. Interestingly, the first radio knot at $\sim$11.8\arcsec\ is  
weakly detected at the X-rays, while no high-energy emission is  
detected from the diffuse radio lobe or the counterlobe. For this  
radio knot we estimated a net count rate of 4$\pm1\times10^{-4}$  
c/s. Using \verb+PIMMS+ and assuming a spectral index $\Gamma=2$, the  
evaluated unabsorbed flux at 0.3--8 keV is 2.3$\times10^{-15}$ \flux\  
corresponding to a flux density at 1 keV of 0.3 nJy. No information  
is present in the optical, because of the smaller field of view of  
\hst.  
  
\subsubsection{\pk}  
  
Figure~\ref{2201in} shows the innermost portions of the jet in \pk. In  
the radio, the jet emerges from the core as a narrow, collimated  
stream ending in a bright enhancement region at 1.1\arcsec\ (knot  
$\alpha$). This is followed by another discrete emission region at  
1.6\arcsec\ (knot $\beta$). Finally, the radio jet flares up again  
into a trident-like structure (A) at 2.2\arcsec, consisting of three  
protruding elongations toward northwest.  
  
In the optical, the emission from knots $\alpha$ and $\beta$ is weak,  
while feature A is bright. The \chandra\ image shows faint and  
continuous X-ray emission from knot $\alpha$ to A. Indeed, there is  
little evidence of a well-defined peak of X-ray emission in the  
longitudinal profiles in Figure~\ref{pklongi}. Only an upper limit to  
the X-ray emission from $\alpha$ is reported in Table~1.  The radio  
and optical peaks of knot $\alpha$ appear slightly offset from each  
other by $\sim$ 0.1\arcsec--0.2\arcsec, a distance that is one order  
of magnitude greater than the \hst\ differential astrometry. However,  
this could likely be an artifact due to galaxy subtraction, as well as  
the optical bump at 0.5\arcsec\ in Figure~\ref{pklongi}. After  
3\arcsec, the optical and X-ray emission decline rapidly into the  
background while at radio frequencies the intensity recovers into a  
smaller knot at 3.8\arcsec. As mentioned in \S~3.2, the double  
structure in the X-ray profile at knot A is probably an artifact of  
the rebinning procedure of the ACIS image.  
  
Figure~\ref{pktras} shows the transverse profiles of the jet at
1.6\arcsec\ (knot $\beta$), 2.2\arcsec\ (knot A), and 2.8\arcsec\
(downstream). The knots are resolved at radio at optical but only
marginally at X-ray wavelengths, with FWHMs $\simeq$ 1.1\arcsec\ at
X-rays.
  
At 4\arcsec\ from the core the radio jet loses collimation, opening in
a wide-cone wind with no discrete emission but rather a continuum, and
changing direction around 5\arcsec\ (Figure~\ref{2201out}). Diffuse
X-ray emission is detected from the inner part of the wind and in
particular in two regions located at 6.2\arcsec\ and 8.2\arcsec\ from
the core. The extracted net count rates are 6$\pm2\times10^{-4}$ c/s
and 7$\pm2\times10^{-4}$ c/s, for the two knots respectively. Also in
this case we used \verb+PIMMS+ to derive the unabsorbed flux at 0.3--8
keV. Assuming a spectral index $\Gamma=2$, we found that
F$_{0.3-8~keV}=3.5\times10^{-15}$ \flux\ and
F$_{0.3-8~keV}=4.1\times10^{-15}$ \flux.

\subsubsection{Inter-knot emission}  
  
As is apparent from Figures~\ref{3clongi} and \ref{pklongi}, the  
emissivity at radio, optical, and X-rays does not go to zero in the  
regions between consecutive knots. The ``dark'' regions between the  
cores and first knots in the optical is an artifact due to the galaxy  
subtraction.   
  
The presence of inter-knot emission is not new, having been noted
previously in several FRII jets (Sambruna et al. 2006 and references
therein). At X-rays, it is still not clear whether the inter-knot
emission is truly continuous or due to unresolved discrete knots. Here
we can only note that this emission is part of the ``spine'' of the
jet. In \S~4.2.1 we will investigate wheather spine and edges have
different spectral properties at X-rays. 
  
\section{X-ray Spectral Analysis} 

\subsection{The Cores} 

Analysis of the core light curves show that the flux did not vary  
within the ACIS exposure of both sources. We thus concentrate on the  
total X-ray spectra, integrated over the entire ACIS exposures, and  
extracted as described in \S~2.2.   
  
The core spectra were fitted with a two-component model, both absorbed  
by Galactic $N_{\rm H}$, including a power law, to represent the AGN,  
and a thermal, to account for the circumnuclear thermal emission  
(\S~3.1).  A component to account for the core pileup, albeit modest,  
was also included. The thermal component was parameterized by the  
model \verb+apec+ in \verb+XSPEC+, with temperatures $kT$ and  
elemental abundances $Z/Z_{\odot}$. During the initial fits we left  
the abundance free to vary from 0.2 and 1. The values of this  
parameter always pegged at one of the limits, thus we fixed the  
abundance at their best fit values, $Z/Z_{\odot}$=1 for \pk\ and 0.2  
for \3c, to obtain more stringent constraints on the remaining  
parameters. For \pk\ the thermal component is not statistically  
significant (P$_F <$ 90\%).  
  
The results of the spectral fits are listed in Table~6, where the  
0.3--8~keV total and power-law fluxes are listed separately. It can be  
inferred from the Table that, as expected from Figure~\ref{core}, the  
thermal component contributes only 2--3\% to the total X-ray emission  
of the core. The core spectrum is rather hard in both sources, $\Gamma  
\sim 1.5$, as in other BL Lacs of similar luminosity (e.g.,  
Donato, Sambruna, \& Gliozzi 2004).

\subsection{The inner jets}  
  
\subsubsection{3C~371}   
  
Spectra were extracted from knots $\beta$ and B, and structure A  
following the procedure described in \S~2.2. In all three cases, the  
0.3--8~keV continua are adequately described by a single power law  
model with Galactic absorption, with the photon indices reported in  
Table~7. These indices are steeper than observed in FRIIs, and similar  
to those derived for FRIs (Worrall et al. 2003). The addition of a  
thermal component is not statistically significant, and a power law is  
preferred over a single thermal model.  
  
Comparing with the photon index of the core, $\Gamma=1.46$ (Table~6),
it is clear that the X-ray continuum steepens going from the
unresolved to the resolved jet. There is also a marginal trend of
gradual softening of the X-ray emission along the jet, however, the
uncertainties are too large to allow firmer conclusions.  As discussed
above, the core PSF contamination to the X-ray emission from knots
$\beta$ and B is negligible because of the choice of the extraction
region for the background.
  
\begin{figure}[ht]  
\begin{center}  
\hbox{  
\centerline{\includegraphics[width=18cm]{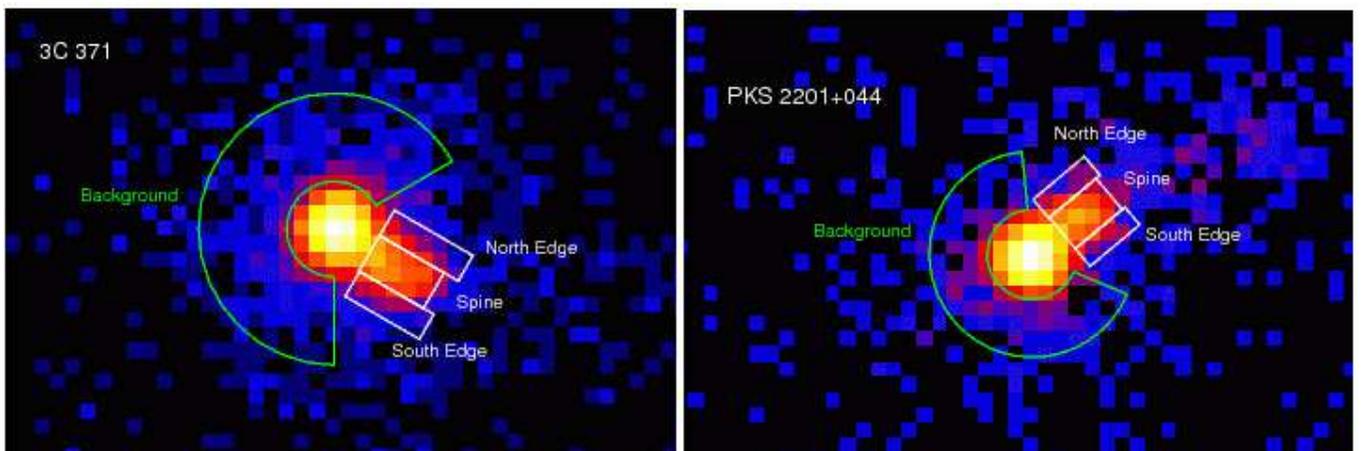}}  
}  
  
\end{center}  
\vspace{-1.0cm}  
\caption{\footnotesize  
{\chandra\ images of the jets of \3c\ and \pk, showing the extraction regions  
for the spine, lateral edges, and background.   
}}  
\label{edges}  
\end{figure}

In \S~3.2.1, evidence was presented that the X-ray jet is resolved
transversally to its axis into a central region, or ``spine'', and
lateral ``edges''. To quantify the X-ray properties of the ``edge''
emission, its ACIS spectrum was extracted. We used rectangular boxes
of length 2.9\arcsec\ and width 0.9\arcsec, located as illustrated in
Figure~\ref{edges}, where the extraction region for the background is
also shown. The X-ray spectra of the two edges were combined in order
to increase the signal-to-noise ratio. Note that the background region
includes an average of the circumnuclear thermal emission and the
wings of the core PSF; these contributions are thus subtracted during
the fits in \verb+XSPEC+. We experimented with the position and size
of the edges but found consistent spectral fits in all cases. We thus
choose the case in Figure~\ref{edges} because it corresponds to the
highest signal-to-noise ratio.
  
Similarly, we extracted the X-ray spectrum of the central portion of
the jet, labeled ``spine'' (Figure~\ref{edges}), using a rectangular
region of dimensions 2.4\arcsec $\times$ 1.3\arcsec. The latter
contains a total of $\sim$1,250 net counts in 0.3--8~keV; this
includes the contribution of $\beta$, B, and A, but also the residual
intra-knot X-ray emission. Because of the limited resolution and
statistics, we make no attempt to remove the knot contributions to the
spectrum.

Note that the extraction region for the ``spine'' is as wide as the
entire X-ray jet; as such, the X-ray emission of both ``spine'' and
``edges'' is blended in the spine spectrum. On the other hand, the
edge regions are sampling the wings of the flux distribution; because
of the lateral extension of the X-ray jet, in the edge regions the
``edges'', if they have a different spectrum, will dominate the
emission depending on the relative intensity of the two components. In
summary, the relative contributions of the ``spine'' and ``edges'',
while blended due to the intrinsic FWHM of ACIS, will differ in the
two extraction regions.

A total of $\sim$260 net counts were extracted from the combined  
edges in 0.3--8~keV, sufficient for spectral analysis. A fit with a  
single power law and Galactic absorption yields a photon index  
$\Gamma_{edges}=2.24^{+0.26}_{-0.25}$, $\chi^2_r=0.91/13$, and an  
observed flux in 2--10~keV of F$_{edges}\sim1.7 \times 10^{-14}$  
\flux. However, the residuals of the single power law model show a  
small ($\sim$ $1\sigma$) flux excess in the energy range 0.8--1~keV; the  
inclusion of a thermal model with temperature $kT\sim 1$ keV, albeit  
statistically not significant ($\Delta\chi^2$=2.7 for 3 additional    
free parameters), indeed produces flatter residuals. The possibility  
of thermal emission along the edges is of particular interest for gas  
entrainment models (Tavecchio et al. 2006). Additionally, structure A  
is highly suggestive of an impact of the jet with external medium, and  
thermal emission of shocked gas would be expected in this case.  
  
For the ``spine'', a fit with a power law yields $\Gamma_{spine}=2.01 \pm  
0.08$, $\chi^2_r=0.74/52$, and 2--10~keV flux F$_{spine} \sim 1.1  
\times 10^{-13}$ \flux. Comparing with the edges, there is a slight  
indication that the spine has a harder X-ray emission than the edges,  
although this is only a $1\sigma$ effect. Better data are needed to
confirm this result. 

To test this result, we used the simulations reported earlier. Using
the same extraction regions as in Figure~\ref{edges} we extracted the
ACIS spectra of the spine and edges. Comparing the two spectra, the
opposite effect is observed: the edges' spectrum is {\it flatter} than
the spine, because clearly dominated by the larger PSF at higher
energies. The difference in slope is still 1$\sigma$. We thus conclude
that there is marginal evidence for a possible spectral difference at
X-rays between the spine and the edges, which needs to be confirmed in
the future in higher-quality data. 

In summary, the X-ray continua of the jet knots, edges, and spine are
well described by a single power law model. We find evidence that the
X-ray emission of the jet is softer than the core.

\subsubsection{PKS~2201+044}   
  
In the case of the discrete jet knots, X-ray spectra were extracted as  
described in \S~2.2, and fitted with a single power law plus Galactic  
absorption model. In all cases the power law is an adequate  
description, with no need for extra (thermal) components. The photon  
indices are similar to those derived for \3c\ (Table~7). There is a  
suggestion for the X-ray continuum of the core to be harder than the  
outer jet (see \S~4.1 and Table~6).   
  
Figure~\ref{edges} shows the extraction regions used for the spine and  
edges of the jet, using similar regions as for \3c. A total of 380 and  
44 counts, respectively, were collected. The 0.3--8~keV continuum of  
the spine is well fitted by a power law with  
$\Gamma_{spine}=2.07\pm0.18$, yielding an observed 2--10~keV flux  
F$_{spine}\sim3.3\times 10^{-14}$ \flux; for the edges, we find  
$\Gamma_{edges}=2.7 \pm 1.8$ and 2--10~keV flux  
F$_{edges}\sim3.0\times 10^{-15}$ \flux.  
  
The conclusions are very similar to \3c. We find an indication that  
the X-ray continuum of the \pk\ jet softens from the core to the  
lobe. However, large uncertainties plague the measurements of the  
slopes, and longer \chandra\ observations are required to confirm the  
spectral gradient along and possibly across the jet.

\subsection{Jet Spectral Energy Distributions}   
  
\begin{figure}[ht]  
\begin{center}  
\hbox{  
\includegraphics[width=7in]{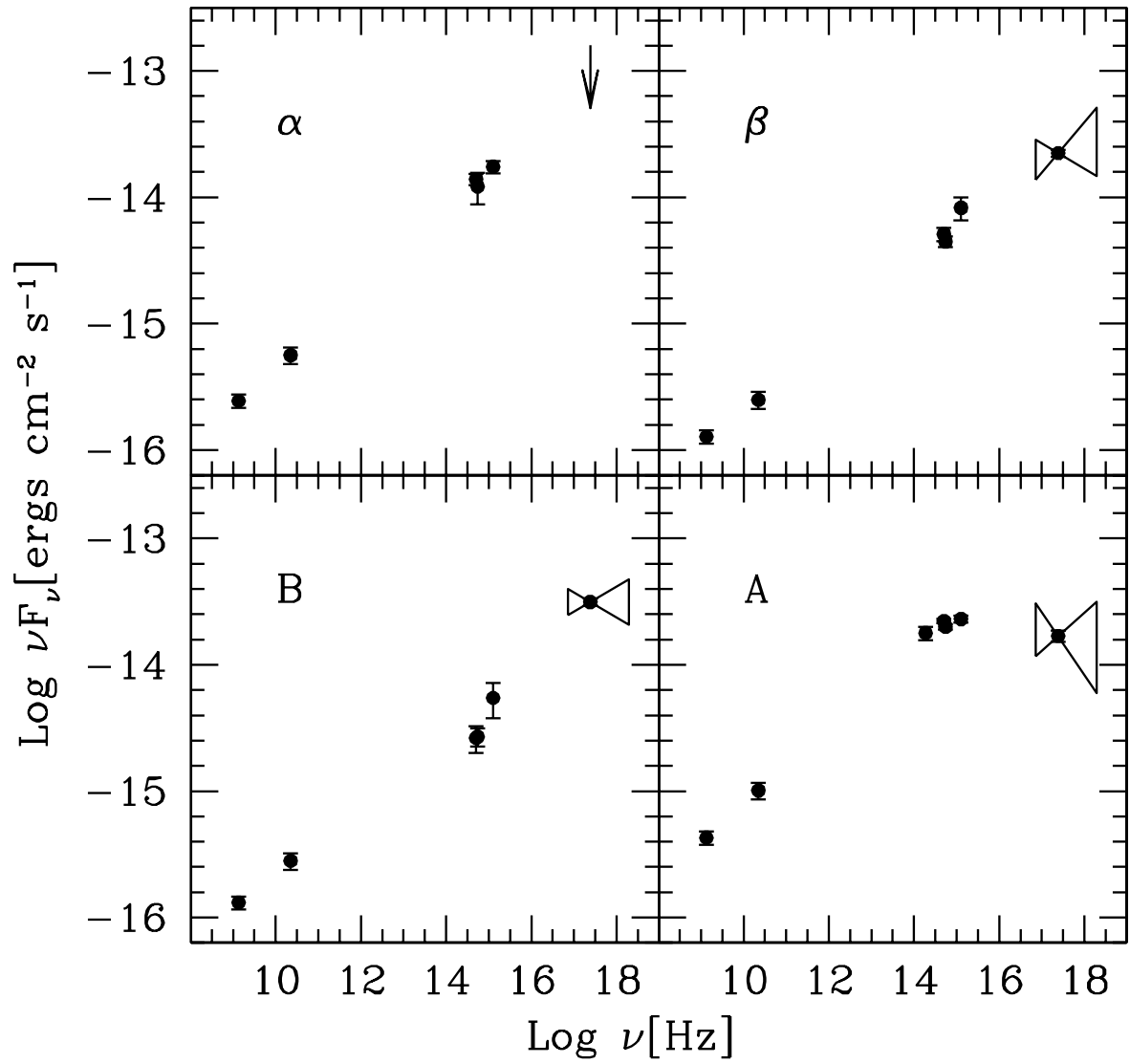}  
}  
\end{center}  
\vspace{-2.0cm}  
\caption{\footnotesize  
{Spectral Energy Distributions of the jet knots for 3C~371.   
}}  
\label{sedknots3c}  
\end{figure}

\begin{figure}[ht]  
\begin{center}  
\hbox{  
\includegraphics[width=7in]{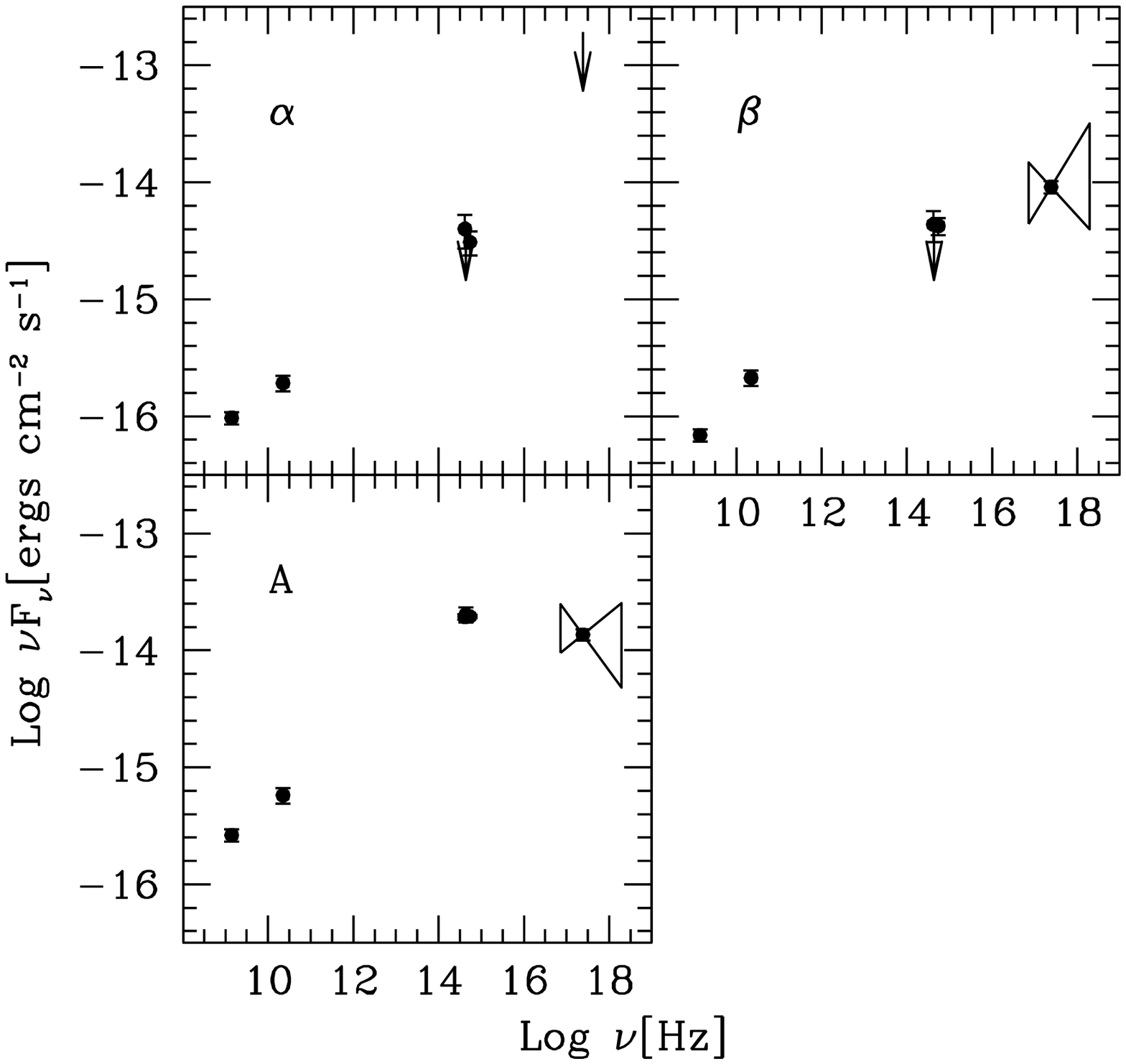}  
}  
\end{center}  
\vspace{-2.0cm}  
\caption{\footnotesize  
{Spectral Energy Distributions of the jet knots for \pk.   
}}  
\label{sedknotspk}  
\end{figure}

The X-ray fluxes of the jet knots were extracted from elliptical  
regions with axes in the range 0.4--1.1\arcsec, as discussed in  
\S~2.2. Because of the higher angular resolution, the optical fluxes  
were extracted using squared regions of smaller sizes, except for
knots A (\S~2.3). The extraction regions at optical and X-rays were
centered on the radio position of the features. The radio extraction
regions were ellipses with axes in the range 0.25--0.4\arcsec,
depending on the size of the knots, but with an area similar to that
covered by the extraction regions at optical. Table~4 lists the
multiwavelength flux densities for the jet features, while the SEDs
are shown in Figures~\ref{sedknots3c} and \ref{sedknotspk}. The fluxes
in the Table and Figures are corrected for absorption, aperture, and
PSF wing contamination (\S~2.2). 

The continuum energy indices are reported in Table~7. In the radio
band the slopes were calculated by interpolating to the corresponding
datapoints; the uncertainties reflect the errors on the fluxes.  For
the radio fluxes, the errors are 12\% at 1.35 and 1.4~GHz, and 15\% at
22.4~GHz. For the optical, the errors on the flux were derived by
averaging the statistical fluctuations of the latter in regions
surrounding the extraction region (\S~2.3).  However, except for knots
A which are the most distant from the cores, uncertainties associated
with the choice of the extraction region are likely to affect the
optical slopes. In \3c, the optical data of knot A clearly indicate a
softening of the optical spectrum with respect to the radio.  The
X-ray slopes were derived from fits to the ACIS spectra of the
individual knots with a power law model (\S~3.3). 
  
Broad-band spectral indices were also calculated, from radio to  
optical ($\alpha_{ro}$), optical to X-rays ($\alpha_{ox}$), and radio  
to X-rays ($\alpha_{rx}$). These indices are listed in Table~7. These  
values were calculated between 5~GHz (using the interpolated flux),  
5550~\AA, and 1~keV.

\section{Discussion}   
  
\subsection{Modeling the knots' SEDs}   
  
The SEDs shown in Figure~\ref{sedknotspk} and the spectral indices
reported in Table~7 reveal some interesting features. The broad-band
indices $\alpha_{ro}$ and $\alpha_{ox}$ are remarkably similar along
the same jet (except for knots A) and between the two jets, as well as
the X-ray slopes $\alpha_X$; the latter, however, have
relatively large errors.  For all knots, the X-ray continuum is softer
than at radio, and softer than the radio to X-ray spectral index; the
exception are knots $\alpha$ for which the X-ray flux could not be
measured. On the whole the SEDs of knots A, in both jets, suggest that
the emission derives from a single synchrotron component following a
power law from the radio to the optical band, peaking close to the
optical wavelength range and softening between the optical and
X-rays. The SEDs of knots B (\3c) and $\beta$ (\pk) are qualitatively
similar to knots A, but with emission peaks in or close to the
X-ray range. The optical fluxes of knot B in \3c\ seem to deviate from
a smooth connection between the radio and X-ray range. However, we note
that this knot has the lowest optical flux and the measurement may
suffer from systematic uncertainties.
 
Thus, the radio-to-X-rays SEDs can be modeled as synchrotron emission
from relativistic electrons with a power law energy distribution up to
some energy $E_b$, above which the power law steepens, probably as an
effect of radiative cooling suffered by the high-energy
electrons. This scenario would be qualitatively consistent with $E_b$
shifting down in energy with increasing distance from the core.
 
\begin{figure}[ht]  
\begin{center}  
\hbox{  
\includegraphics[width=17cm]{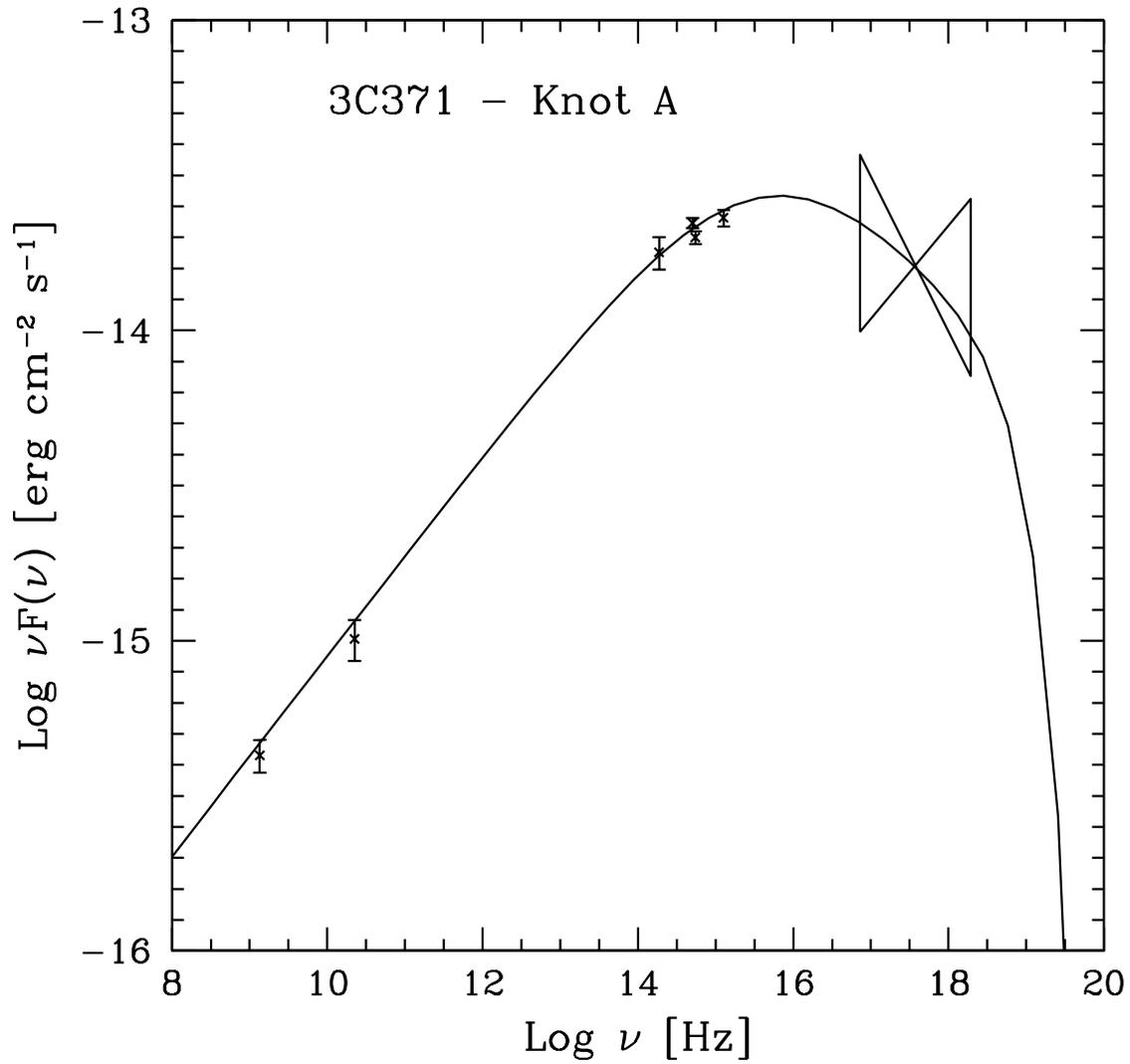}  
}  
\end{center}  
\vspace{-2.0cm}  
\caption{\footnotesize  
{Spectral Energy Distributions of the knot A of the \3c\ jet. The
lines are fits to the data with a model assuming synchrotron emission
from a broken power law (solid line) distribution of electrons.  }}
\label{sedknota3c}  
\end{figure}

As an example we model the overall emission observed from knot A of
\3c\ (Figure~\ref{sedknota3c}); given the similarity of the SEDs, the
modeling of the emission from the other knots involves very similar
parameter. We assume that the SED is described by synchrotron
emission from a broken power law electron energy distribution. The
emitting volume, assuming a sphere with radius of 0.3\arcsec, is
$V=4\times 10^{63}$ cm$^3$ (equivalent to 0.14 kpc$^3$), the magnetic
field (assumed to be in equipartition with the electrons) is $B=73$
$\mu$G, the Doppler factor is fixed to $\delta =2$ and the electron
normalization is $K=1.5\times 10^{-4}$ part/cm$^3$. The distribution
extends from minimum Lorentz factor $\gamma_{\rm min}=1$ to the
maximum $\gamma_{\rm max}=10^8$ and we assume that the break is
located at the Lorentz factor $\gamma_{\rm br}=3\times 10^6$. The
low-energy slope is $n_1=2.35$, while the high-energy slope has been
fixed to $n_2=3.55$ (close to the value $n_2=n_1+1$ expected if the
break derives from radiative cooling of the high-energy
electrons). Broken power-law distributions are commonly adopted to
describe the multifrequency data in other FRI jets (see e.g.,
Birkinshaw et al. 2002, Hardcastle et al. 2002).
 
Even assuming equipartition between magnetic and particle energies,  
the model is not sufficiently constrained by the observed SEDs, thus 
several model parameters have been chosen plausibly but arbitrarily. 
In particular, we lack information on the low-energy end of the electron  
distribution, $\gamma_{\rm min}$, and on the value of the Doppler factor,  
$\delta$, both entering in the determination of the equipartition magnetic  
field.
 
\begin{figure}[ht]  
\begin{center}  
\hbox{  
\centerline{\includegraphics[width=10cm]{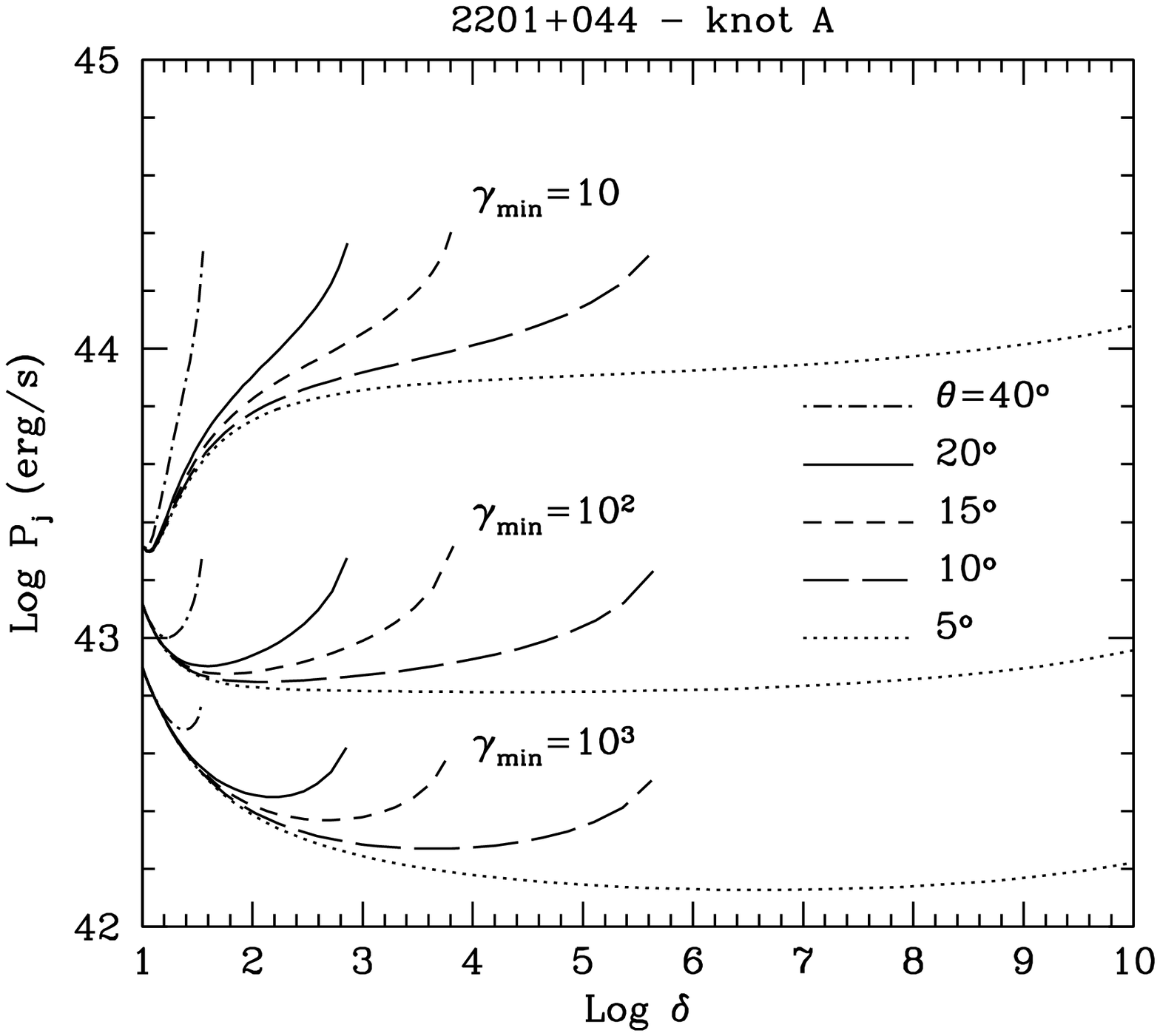}}  
}  
  
\vspace{-1.0cm}  
  
\hbox{  
\centerline{\includegraphics[width=10cm]{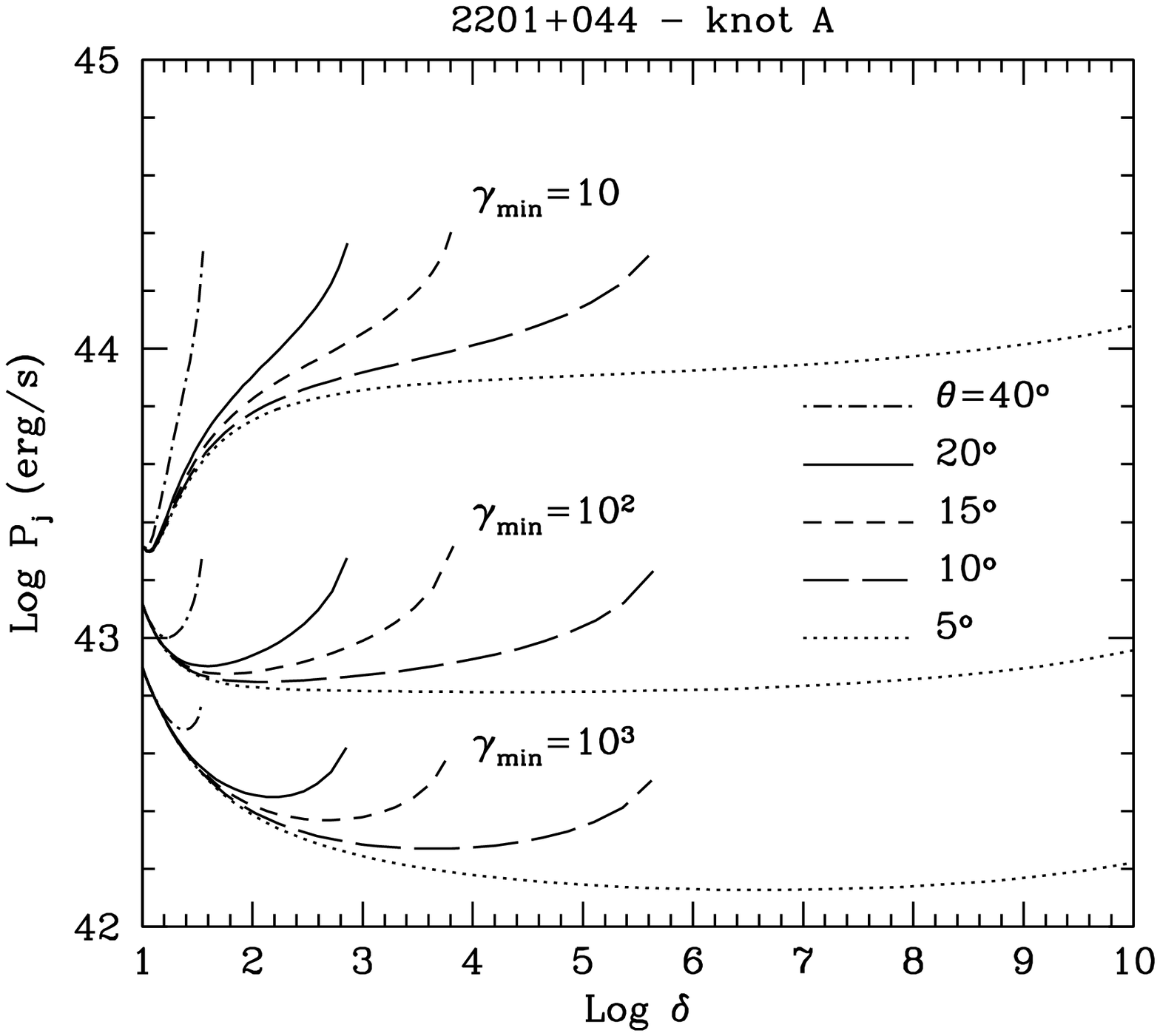}}  
}  
\end{center}  
\vspace{-2.0cm}  
\caption{
{Jet kinetic power as a function of the Doppler factor calculated at knots A  
of \3c\ (top) and \pk\ (bottom).  The different curves show the power for  
different values of the minimum Lorentz factor of the relativistic  
electron, $\gamma_{\rm min}$ and for different jet inclinations (see  
the text for more details).}}  
\label{power-fab}  
\end{figure}

In order to explore the effect of a different parameter choice on the
estimate of the basic physical quantities of the jet we use analytical
expressions relating the observed radio spectrum to the physical
parameters (e.g., Tavecchio et al. 2006). The results are reported in
Figure~\ref{power-fab}, both for knots A of \3c\ and \pk.
 
For a given choice of the values of the Doppler factor $\delta$ and
$\gamma_{\rm min}$, we calculate the values of the other parameters
($B$ and $K$) from the condition of equipartition and the value
of the observed radio flux. In turn, the value of the Doppler factor
can be translated into a value of the bulk Lorentz factor of the flow,
$\Gamma$, necessary to calculate the jet power, assuming a given
viewing angle, $\theta$. 

In the plots we show the resulting value of the power carried by the
jet\footnote{We use the expression of the jet power reported in
Schwartz et al. (2006).} as a function of the Doppler factor, for
various choices of minimum Lorentz factor ($\gamma_{\rm min}=10, 100,
10^3$) and viewing angles. The latter assumes the following values:
$\theta =18$\deg, 15\deg, 10\deg, and 5\deg\ for \3c, for which the
upper limit comes from the observed jet/counterjet ratio measured at
VLBI scales (Gomez \& Marscher 2000); $\theta$=40\deg, 20\deg, 15\deg,
10\deg, and 5\deg\ for \pk, whose upper limit is obtained assuming a
correlation between the observed radio power of the core and the total
power, expected from the relativistic boosting at the base of the jet,
at low frequency (Giroletti et al. 2004). For both sources, for
magnetic fields $B<100$ $\mu$G, the derived power ranges from $P_{\rm
j}=10^{42}$ \lum\ to $P_{\rm j}=10^{45}$ \lum, consistent with the
power inferred for other FRI jets (e.g., Bicknell \& Begelman 1996; 
Laing \& Bridle 2004). Note that, for a given choice of $\theta$ there
is a limited range of allowed values for $\delta$: this fact
translates into the limited range of the parameters covered by the
lines in Figure~\ref{power-fab}. The contribution of Inverse Compton
emission, both SSC and External, is negligible at these frequencies
for all the choices of the parameters.

An interesting issue concerns the nature of the knots observed in the  
two sources. A careful inspection of the multifrequency maps of both  
sources suggests a difference between the inner knots ($\alpha$,  
$\beta$, and B for \3c; $\alpha$ and $\beta$ for \pk) and knots  
A. While the first features show a  compact morphology, knots A,  
instead, are resolved at all wavelengths. In the case of knot A of  
\3c, moreover, the emission is more concentrated on one side of the  
jet (see Figure~\ref{3c371in}). As already noted (Nilsson et al. 1997),  
the position of this knot coincides with the location of a moderate  
change in the direction of the jet, clearly visible from the maps in  
Figure~\ref{3c371out}. It is tempting to associate the strong emission   
from knot A with the presence of an internal shock induced into the   
supersonical flow by the bending experienced by the jet. On the other   
hand, the location of knot A in \pk\ is close to the point where the jet   
seems to loose collimation and the surface brightness starts to decline.   
In this sense, this knot shares some similarities with the ``flaring   
points'' frequently observed in FRI jets, thought to mark the   
recollimation shock(s) expected when the decreasing pressure of the   
expanding jet reaches the value of the external medium (e.g., the   
discussion in Laing \& Bridle 2002).  

\subsection{The jets of \3c\ and \pk\ in a broader context}  
  
This paper presented multiwavelength observations with \chandra, \hst,  
and the \vla\ of the relativistic jets in the nearby radio galaxies  
\3c\ and \pk. While \chandra\ has discovered and studied many extragalactic  
jets in FRIs and FRIIs so far (a list is maintained at XJET
Website\footnote[2]{http://hea-www.harvard.edu/XJET/}), the sources of
this paper stand out because of their ``intermediate'' properties.
 
From the point of view of their cores, both sources are classified as
BL Lacs based on the continuum variability and polarization. However,
narrow and broad emission lines were at times detected from the cores
leading to an earlier classification as N galaxy (\3c) and Seyfert~1
(\pk). In \3c, however, the [OIII] emission appears extended in a S-W
direction toward a nearby companion galaxy (Stickel, Fried,
\& K\"uhr 1993). While the core-jet radio morphology resembles that of  
FRIIs, particularly in the case of \3c\ which shows prominent lobes,  
the total radio power and optical magnitude of the hosts of \3c\ and  
\pk\ are similar to other FRIs (Ledlow \& Owen 1996). 
  
The data presented here support the idea that \3c\ and \pk\ are
intermediate sources between FRIs and FRIIs. As in FRIs, the X-ray
emission peaks closer to the nucleus while the longer wavelength
emission brightens towards the end. The polarization structure shows
prominent features at the so-called ``flaring point'' on a similar
physical distance from the nucleus as in other FRIs (e.g., Laing \&
Bridle 2002). In fact, the properties of \3c\ and \pk\ resemble those
of PKS~0521--365, which also exhibits an X-ray jet (Birkinshaw,
Worrall, \& Hardcastle 2002). In the latter source, a knot of X-ray
emission was detected in earlier \chandra\ observations at 2\arcsec\
from the nucleus, while the radio jet is three times longer. As in the
sources of this paper, PKS~0521--365 is classified as a BL Lac
although narrow and broad optical lines were detected from the core
(Danziger et al. 1979).  

On the other hand, the X-ray jets are one-sided as in FRIIs, but,
unlike in FRIIs, they appear to be less collimated and do not exhibit
the typical hard X-ray spectra present in high-power radio
sources. The jet powers are also lower than typically derived for
FRIIs (Sambruna et al. 2004). This is in agreement with a unified
scenario where the differences can be understood on a physical
basis. The different radio morphologies could result from jet
deceleration occurring within the galaxy (FRIs) or outside the galaxy
(FRIIs). Moreover, if the deceleration mechanism is entrainment of
external gas, the scale length for deceleration is related to the jet
power, which naturally accounts for the correlation of radio power and
morphology (Bicknell 1995; Tavecchio et al. 2006).
 
Based on the multiwavelength properties of the jets one would conclude  
that these three sources - \3c, \pk, and PKS~0521--365 - are FRIs with  
favorable orientations of their jets, i.e., more beamed than classical  
FRIs. The presence of broad optical emission lines from at  
least one of their nuclei may question this hypothesis, as FRIs  
generally are inefficient at producing broad optical lines (Baum,  
Zirbel, \& O'Dea 1995). On the other hand, in higher luminosity BL  
Lacs, usually associated with FRIs, low states of the non-thermal  
continuum often uncover broad emission lines (Pian et al. 2002;  
Vermeulen et al. 1995).

\section{Summary}   
  
In this paper we presented and discussed multiwavelength observations  
of the jets in the nearby radio galaxies \3c\ and \pk, acquired using  
\chandra, \hst, \vla, and \merlin. Our primary findings are:   
  
\begin{itemize}   
  
\item The morphology of the jets is similar, with the X-ray emission  
brightening close to the nucleus and fading thereafter. This is  
comparable with other FRIs detected with \chandra;   
  
\item The SEDs of the knots are consistent with a single emission  
component from radio to X-rays, interpreted as synchrotron. The   
X-ray continuum steepens from the core to the end of the jet;   
  
\item The knots are resolved in a direction perpendicular to the jet  
axis at radio, optical, and X-rays. Radio polarization images indicate  
the presence of an edge-spine structure in both jets; 
  
\item The derived jet power ranges from $P_{\rm  
j}=10^{42}$ \lum\ to $P_{\rm j}=10^{45}$ \lum;
  
\item Overall, the X-ray jets of \3c\ and \pk\ exhibit intermediate
properties between low- and high-power radio sources. 
  
\end{itemize}   
  
While the jet and circumnuclear properties are completely consistent  
with those of other FRIs studied with \chandra, the nuclear properties  
of \3c\ and \pk\ are unusual, in showing narrow and broad optical  
lines. It is thus possible we are seeing these jets at closer angles  
than in classical FRIs.

\acknowledgements  
  
Many thanks to the anonymous referee for constructive criticism and
a generally  supportive attitude. The \vla\ is a facility of the National
Radio Astronomy Observatory is operated by Associated Universities,
Inc. under a cooperative agreement with the National Science
Foundation. \merlin\ is a UK National Facility operated by the
University of Manchester at Jodrell Bank Observatory on behalf of
PPARC. We thank Dr. Anita Richards for performing the initial pipeline
calibration of the \merlin\ data.  Based in part on observations made
with the NASA/ESA Hubble Space Telescope, obtained from the data
archive at the STScI.  STScI is operated by the Association of
Universities for Research in Astronomy, Inc. under NASA contract NAS
5-26555.


\clearpage  

\begin{table}
\caption{Observed Properties of X-ray jet knots\label{jetX_01}}
\begin{center}
\begin{tabular}{l c c c c c} 
\multicolumn{6}{l}{   } \\ \hline
~~Knot & Distance & Counts Rate & Backg & F$_{\rm 1~keV}$ & F$_{\rm 0.3-8 keV}$  \\
~~~~(1)    &  (2)   &  (3)     &  (4)   &  (5) & (6)      \\ \hline        
\multicolumn{6}{c}{3C~371} \\ \hline 
$\alpha$    & 0.8 & $<$32.2     & $\cdots$  & $<$66.23  & $<$80.12  \\
$\beta$     & 1.3 & 6.2$\pm$0.4 & 1.1 &   9.23  &    7.35  \\
B           & 1.9 & 5.9$\pm$0.4 & 0.4 &   12.98  &    10.26  \\
A           & 3.1 & 2.4$\pm$0.3 & 0.1 &    6.98  &     4.82  \\ \hline 
\multicolumn{6}{c}{PKS~2201+044} \\ \hline 
$\alpha$   & 1.1 & $<$29.7     & $\cdots$ &  $<$79.20  & $<$81.18  \\
$\beta$    & 1.6 & 1.8$\pm$0.3 & 0.5 &    3.77  &     3.10  \\
A        & 2.2 & 2.0$\pm$0.2 & 0.1 &    5.63  &     4.21  \\ \hline
                      
\end{tabular}
\end{center}
\tablecomments{{\bf Columns explanation}: 1=Knot name; 2=Distance from the core (in arcsec); 
3=Net count rate in the 0.3--8 keV band in units of $10^{-3}$ c/s;
4=Background count rate rescaled to the knot extraction region size in
the 0.3--8 keV band in units of $10^{-3}$ c/s; 5=X-ray flux density at
1 keV (in nJy); 6=X-ray flux in the 0.3--8 keV band in units of
$10^{-14}$ \flux. Fluxes and flux densities are aperture-corrected,
background subtracted, and corrected for the contamination of the PSF
wings of nearby knots. The total correction factors are 1.6, 2.3, and
2.7 for knots $\beta$, B, and A of 3C~371; and 2.2 and 2.9 for knots
$\beta$ and A of PKS~2201+044. The upper limits are 3$\sigma$ above the
surrounding background.}
\end{table}



\begin{table}
\caption{HST Observation Log\label{loghst}}
\begin{center}
\begin{tabular}{l l l c c c} 
\multicolumn{6}{l}{   } \\ \hline
Instrument &  Aperture  &  Filter    & Wavelength & Obs. Date & Exp. Time \\
~~~~(1)    &  ~~~~(2)   &  ~~(3)     &  (4)~~~~   &  (5)     &   (6)     \\
\hline        
\multicolumn{6}{c}{3C~371} \\
\hline        
NICMOS & NIC2-FIX  &  F160W  & 16030 & 1997-06-01 & 1152 \\ 
WFPC2  & PC1-FIX   &  F702W  &  6917 & 1994-08-18 &  560 \\
ACS    & WFC1-2K   &  F606W  &  5918 & 2003-09-20 &  658 \\
WFPC2  & PC1       &  F555W  &  5443 & 2001-06-23 & 3640 \\
STIS   & F25QTZ    &  MIRNUV &  2357 & 2000-09-21 & 7989 \\
\hline        
\multicolumn{6}{c}{PKS~2201+044} \\
\hline        
WFPC2  & PC1-FIX   &  F702W  &  6917 & 1998-12-13 &  610 \\
WFPC2  & PC1       &  F555W  &  5443 & 2001-07-16 & 4240 \\
STIS   & F28X50LP  &  MIRVIS &  7218 & 2001-07-16 & 2536 \\
\hline
                      
\end{tabular}
\end{center}

\tablecomments{{\bf Columns explanation}: 1=Instrument; 2=Aperture; 3=Filter; 
4=Pivot wavelength (\AA); 5=Observation date (yyyy-mm-dd); 
6=Calculated total exposure time (s).}
\end{table}

  
\begin{table}
\small
\caption{Radio Observation Log\label{radiolog}}
\begin{center}
\begin{tabular}{l l c c c c c}
\multicolumn{7}{l}{   } \\ \hline
Telescope & Obs. Date & Frequency & Exp. Time  & Resolution  & Program & Obs./Ref.\\
~~~~(1)   &  ~~~~(2)  &   (3)     &  (4)       &  (5)        &   (6)   & (7)      \\ \hline
\multicolumn{7}{c}{3C~371} \\ \hline        
MERLIN   & 1998 Apr 1/2   &  1.35 & 18    & 0.177x0.149  &  --       & (1) \\
VLA    
         & 1986 Mar 21    &  4.86 &  0.79 & 0.804x0.387  & AC150    & (2) \\
         & 1985 May 05    &  4.99 &       &   2.1x2.1    &  --      & (3) \\
         & 2002 Jul 21/22 & 22.46 &  3.31 &   0.2x0.2    & AC641    & (4) \\ \hline   
\multicolumn{7}{c}{PKS~2201+044} \\ \hline        		    
MERLIN   & 2000 Dec 21    &  1.41 & 11    & 0.336x0.169 & MN/01A/04 & (4) \\
VLA 
         & 2002 Jun 06    &  8.46 &  1.99 &  0.75x0.75  & AC641     & (4) \\
         & 2002 May 28    & 22.46 &  3.3  & 0.252x0.124 & AC641     & (4) \\ \hline
                      
\end{tabular}
\end{center}

\tablecomments{{\bf Columns explanation}: 1=Telescope; 2=Observation Date;
3=Frequency (in GHz); 4=Exposure Time (in hours); 
5=Resolution (in arcsec); 6=Program Code; 7=Observer or Published reference:
(1)=Pesce et al. (2001); (2)=J. Conway; (3)=Wrobel \& Lind (1990); (4)=C.C. Cheung.}
\end{table}


\begin{table}
\small
\caption{Multiwavelength fluxes\label{multiwave}}
\begin{center}
\begin{tabular}{l c c c c c c c}
\multicolumn{6}{l}{   } \\ \hline
& & & & & & &  \\
\multicolumn{8}{c}{3C~371} \\
& & & & & & &  \\
\hline        
~~Knot & Distance & F$_{\rm 1~keV}$ & F$_{\rm 2358~\AA}$ & F$_{\rm 5443~\AA}$ & F$_{\rm 5918~\AA}$ & F$_{\rm 22.46~GHz}$ & F$_{\rm 1.35~GHz}$  \\
~~~~(1)    &  (2)   &  (3)     &  (4)   &  (5)     &   (6)  &  (7)     &  (8)     \\
\hline        
$\alpha$     & 0.9 & $<$66.2 & 1.4 & 2.2 & 2.7 & 2.5 & 18.1  \\
$\beta$      & 1.3 &    9.2 & 0.7 & 0.8 & 1.0 & 1.1 &  9.4  \\
B        & 1.8 &    13.0 & 0.4 & 0.5 & 0.5 & 1.3 &  9.7  \\
A\tablenotemark{a}  & 3.1 &     7.0 & 1.8 & 3.6 & 4.4 & 4.5 & 31.7  \\  
\hline        
& & & & & & &  \\
\multicolumn{8}{c}{PKS~2201+044} \\
& & & & & & &  \\
\hline        
~~Knot & Distance & F$_{\rm 1~keV}$ & F$_{\rm 5443~\AA}$ & F$_{\rm 6917~\AA}$ & F$_{\rm 7219~\AA}$ & F$_{\rm 22.46~GHz}$ & F$_{\rm 1.41~GHz}$ \\
~~~~(1)    &  (2)   &  (3)     &  (4)   &  (5)     &   (6)  &  (7)     &  (8)     \\
\hline        
$\alpha$   & 1.1 & $<$79.2  & 0.6 & $<$1.1 & 1.0 & 0.9 &  6.9 \\ 
$\beta$  & 1.6 &     3.8  & 0.8 & $<$1.1 & 1.0 & 1.0 &  4.9 \\ 
A        & 2.2 &     5.6  & 3.5 &  4.7 & 4.6 & 2.6 & 18.7 \\
\hline
                      
\end{tabular}
\end{center}
\tablecomments{{\bf Columns explanation}: 1=Knot name; 2=Distance from the core (in arcsec); 
3=X-ray flux density at 1 keV (in nJy); 4-6=Optical flux densities at the 
indicated wavelengths (in $\mu$Jy). X-ray and optical fluxes are 
absorption- and aperture-corrected. The upper limits are 3$\sigma$; 7-8=Radio flux densities at the indicated frequencies (in mJy).}  

\tablenotetext{a}{This feature was also detected by NICMOS at 
$\lambda=16030$~\AA, with F$_{\rm 16030~\AA}=9.5 \mu$Jy 
after galaxy and diffraction spike subtraction, and correcting for aperture.}
\end{table}

  

\begin{table}
\caption{Core Radial Profiles\label{spatial}}
\begin{center}
\begin{tabular}{l l l} 
\multicolumn{3}{l}{   } \\ \hline
                         &      3C~371             &   PKS~2201+044  \\
\hline
r$_c$                    & 5.05$^{+5.68}_{-5.05}$  &  17.38$\pm$7.35 \\
$\beta$                  &    0.42$\pm$0.08        &       0.67 fix    \\
Norm.                    & 1.81$^{+1.90}_{-1.81}$  &  0.37$\pm$0.22  \\
$\chi^2_{\rm r}$/d.o.f.  &     1.98/101            &     1.61/102    \\
P$_F$                    &      $>$99.99\%         &      99.1\%     \\
\hline                                                             

\end{tabular}
\end{center}
\tablecomments{Results of the spatial analysis of the radial profile
around the cores of 3C~371 and PKS~2201+044. We report the core radius
(in arcsec), $\beta$, and normalization (in units of $10^{-5}$ counts
s$^{-1}$ arcsec$^{-2}$) of the $\beta$ model, the reduced $\chi^2$ and
degrees of freedom, and F-test probability P$_F$ for the addition of the
$\beta$ model to the fit.}
\end{table}

  

\begin{table}
\caption{Core Spectral Fits\label{spectral}}
\begin{center}
\begin{tabular}{l l  c c } 
\multicolumn{4}{l}{   } \\ \hline
    &    &    3C~371              &   PKS~2201+044     \\
\hline

Thermal                 & kT        & 0.29$\pm$0.04  &        \\
                        & Z         & 0.2 fix         &                  \\
Power law               & $\Gamma$  & 1.46$\pm$0.06  & 1.59$\pm$0.04       \\
$\chi^2_{\rm r}$/d.o.f. &           & 0.94/176       & 1.02/110            \\
\hline
Total Flux              &           & 3.4           &                 \\
Total Lum.              &           & 20.1          &                 \\
P.law Flux                &           & 3.3           & 1.41                \\ 
P.law Lum.                &           & 19.6          & 2.61                \\
\hline   

\end{tabular}
\end{center}
\tablecomments{Results of the spectral analysis of the cores of 3C~371
and PKS~2201+044. The temperature is in keV and the abundance $Z$ is with
respect to solar values, $Z/Z_{\odot}$. The observed fluxes and
intrinsic (absorption-corrected) luminosities are in the energy range
0.3--8~keV and in units of $10^{-12}$ \flux\ and $10^{42}$ \lum,
respectively.}
\end{table}

  

\begin{table}
\caption{Jet Spectral Indices\label{jetX_02}}
\begin{center}
\begin{tabular}{l c c c c ccc } 
\multicolumn{8}{l}{   } \\ \hline
Source & Knot & $\alpha_{rx}$ & $\alpha_{ro}$ & $\alpha_{ox}$ & $\alpha_R$ &  $\alpha_O$ & $\alpha_X$  \\  \hline 
3C~371 & $\alpha$ & $>$0.66 & 0.70 & $>$0.58 & 0.70$\pm$0.07 & 0.74 $\pm$ 0.17 & $\cdots$  \\
       & $\beta$  & 0.73 & 0.72 & 0.74 & 0.76$\pm$0.07 & 0.39 $\pm$ 0.25 & 0.9 $\pm$ 0.3 \\ 
       & B        & 0.71 & 0.77 & 0.60 & 0.73$\pm$0.07 & 0.19 $\pm$ 0.39 & 1.0 $\pm$ 0.2  \\
       & A        & 0.82 & 0.70 & 1.03 & 0.69 $\pm$ 0.07 & 0.92 $\pm $ 0.08 & 1.1 $\pm$ 0.4 \\ \hline
& & & & & & & \\
PKS 2201+044 & $\alpha$ & $>$0.59 & $>$0.73 & $>$0.32 & 0.75$\pm$0.07 & 1.93 $\pm$ 1.40 & $\cdots$ \\
             & $\beta$  & 0.75 & $>$0.69 & $>$0.88 & 0.59$\pm$0.07 & 1.08 $\pm$ 1.20 & 0.9 $\pm$ 0.5\\
             & A        & 0.80 & 0.66 & 1.06 & 0.71 $\pm$ 0.07 &  0.97 $\pm$ 0.20 & 1.1 $\pm$ 0.4 \\ \hline 
                      
\end{tabular}
\end{center}
\tablecomments{The broad-band indices were calculated using the monochromatic fluxes at 
5~GHz, 5550\AA, and 1~keV (see text and Table~4).}
\end{table}


\clearpage

  
\vspace{-1.0cm}  
  


\vspace{-1.0cm}  
  

\vspace{-1.0cm}  
  
  
\vspace{-1.0cm}  

  

\vspace{-1.0cm}  
  

  
\vspace{-1.0cm}  


\vspace{-1.0cm}  
  


\vspace{-1.0cm}  


\vspace{-1.0cm}  

  
\vspace{-1.0cm}  

  
\vspace{-1.0cm}  

  
\vspace{-1.0cm}  

  
\vspace{-1.0cm}  
     
\end{document}